# Excess resistivity in graphene superlattices caused by umklapp electron-electron scattering

J. R. Wallbank[1], R. Krishna Kumar[1,2] , M. Holwill[1,2], Z. Wang[2], G. H. Auton[1], J. Birkbeck[1,2], A. Mishchenko[1,2], L. A. Ponomarenko[3], K. Watanabe[4], T. Taniguchi[4], K. S. Novoselov[1,2], I. L. Aleiner[5], A. K. Geim[1,2], V. I. Fal'ko[1,2]

[1]National Graphene Institute, University of Manchester, Manchester M13 9PL, UK

[2]School of Physics & Astronomy, University of Manchester, Oxford Road, Manchester M13 9PL, UK

[3]Department of Physics, Lancaster University, Lancaster LA1 4YW, United Kingdom

[4]National Institute for Materials Science, 1-1 Namiki, Tsukuba 305-0044, Japan

[5]Physics Department, Columbia University, New York, NY 10027, USA

**Umklapp processes play a fundamental role as the only intrinsic mechanism that allows electrons to transfer momentum to the crystal lattice and, therefore, provide a finite electrical resistance in pure metals[1,2]. However, umklapp scattering has proven to be elusive in experiment as it is easily obscured by other dissipation mechanisms[1,2,3,4,5,6]. Here we show that electron-electron umklapp scattering dominates the transport properties of graphene-on-boron-nitride superlattices over a wide range of temperatures and carrier densities. The umklapp processes cause giant excess resistivity that rapidly increases with increasing the superlattice period and are responsible for deterioration of the room-temperature mobility by more than an order of magnitude as compared to standard, non-superlattice graphene devices. The umklapp scattering exhibits a quadratic temperature dependence accompanied by a pronounced electron-hole asymmetry with the effect being much stronger for holes rather than electrons. Aside from fundamental interest, our results have direct implications for design of possible electronic devices based on heterostructures featuring superlattices.**

In umklapp electron-electron (Uee) scattering, a crystal lattice gives a pair of interacting electrons a momentum kick such that

$$\vec{k}_1 + \vec{k}_2 = \vec{k}_3 + \vec{k}_4 + \vec{g}\,, \tag{1}$$

where $\hbar\vec{k}_{1,2}$ and $\hbar\vec{k}_{3,4}$ are the initial and final momenta of the two electrons near the Fermi level, respectively, and $\vec{g}$ is a non-zero reciprocal lattice vector of the crystal. In clean metals, normal electron-electron (e-e) scattering, such that $\vec{g} = 0$, does not lead to a finite resistance because e-e collisions do not relax the momentum imparted to the electron system by the electric field (unless the charge carriers involved have the opposite polarity so that, e.g., electrons scatter at thermally excited holes[7,8,9]). This can be understood by considering the case of head-on collisions along the direction of the electric field: If one electron is scattered backwards, the other must be scattered in the forward direction in order to conserve momentum, as illustrated in Fig. 1a (left) for Dirac electrons in one of the graphene valleys. In contrast, in umklapp processes (Fig. 1a, right), both electrons near the Fermi level can be scattered in the backward direction with the Bragg momentum, $\hbar\vec{g}$, transferred to the lattice. This behavior originates from the peculiar nature of electrons in periodic potentials, whose momentum is conserved only up to one reciprocal lattice vector ($\hbar\vec{g}$).

Although recent theories predict a dominant role of Uee processes in determining the high-temperature ($T$) resistivity for some classes of conductors[10,11], experimental evidence for Uee scattering has so far been reported only for a few ultraclean metals[1,2] and in laterally modulated two-dimensional electron gasses in GaAs/AlGaAs quantum wells[3,4,5,6]. In both cases the umklapp contribution was relatively small and noticeable only at $T < 15$ K, being dwarfed by other thermal processes at higher $T$. In this report, we show both theoretically and experimentally that, in graphene moiré superlattices, Uee scattering dominates $T$-dependent resistivity over a wide range of carrier densities, $n$ (a representative miniband[12,13,14] for Dirac electrons in graphene superlattices is shown in Fig. 1a, right).

The studied devices (inset of Fig. 1b) were fabricated using the standard methods for assembling encapsulated graphene/hexagonal boron-nitride (hBN) devices (Supplementary Section 1) where a superlattice (SL) was engineered by aligning graphene with an hBN substrate. This produces a moiré pattern[15,16,17,18] due to the small lattice mismatch ($\delta \approx 1.8$ %) between the two crystals (inset of Fig. 1c), which in turn creates a superlattice potential with a period of around 15 nm for perfect alignment. The SL potential acts on graphene's charge carriers and causes significant reconstruction of its electronic spectrum. In particular, a mini-Brillouin zone is created around the Dirac points of graphene[13,14], whose size is determined by the misalignment angle, $\theta$, and resulting moiré period, $\lambda$. Because the Brillouin zone is small as compared to normal metals, Uee scattering becomes dominant in graphene/hBN superlattices. We present our results referring to 6 superlattice devices and, for comparison, a reference device in which the graphene and hBN axes were intentionally misaligned ($\theta > 15°$; $\lambda < 3$ nm). For aligned samples, $\lambda$ was determined from the frequency of Brown-Zak oscillations using magnetotransport measurements[19] (Supplementary

Section 2). All our devices exhibited high mobility's of up to 500,000 $cm^2$/Vs at liquid helium $T$ and low doping but, at elevated $T$, their resistive behavior strongly depended on λ (see below).

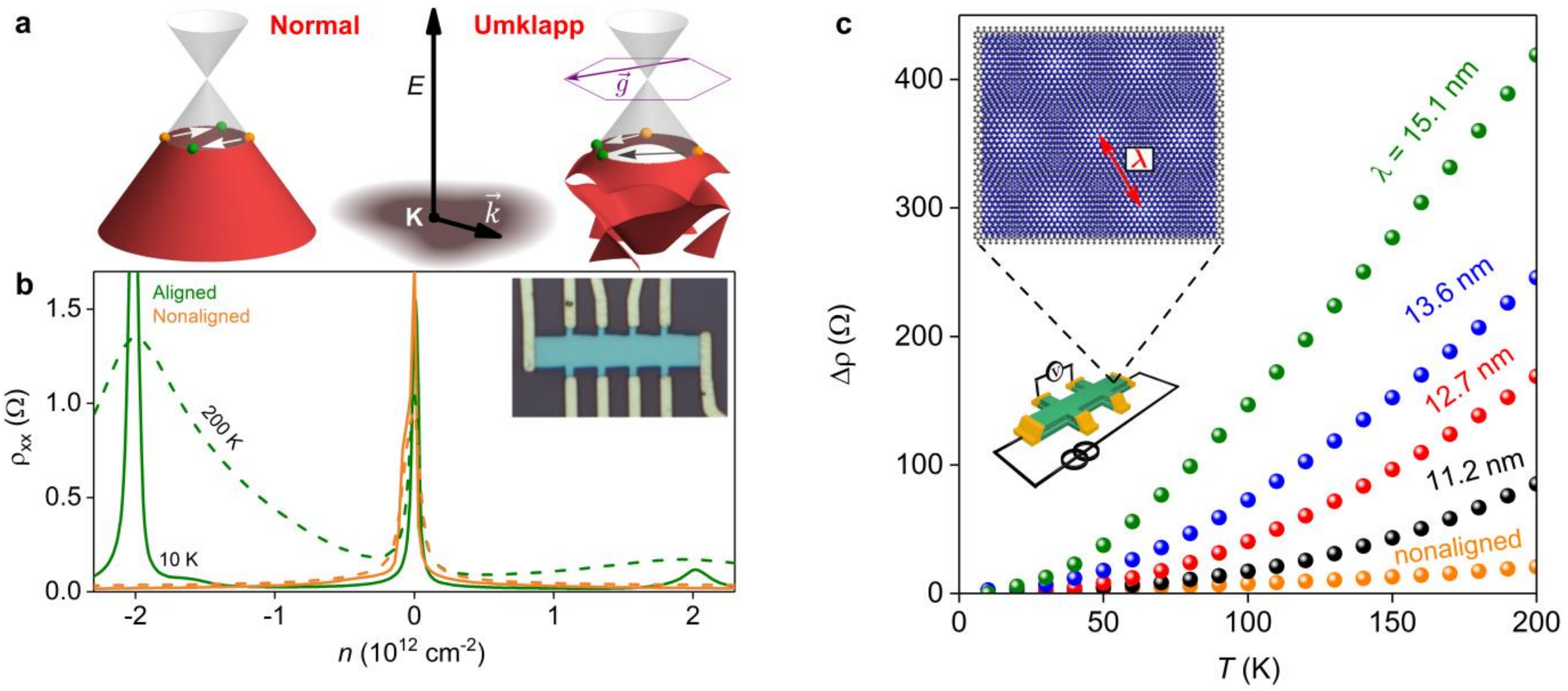

**Figure 1| Umklapp scattering and excess resistivity in graphene superlattices. a,** Normal e-e scattering for, e.g., holes in graphene does not lead to resistivity (left), in contrast to the umklapp scattering for holes in a graphene superlattice (right). Here we also illustrate the SL Brillouin zone (purple hexagon) and SL minibands in the valence band of graphene. **b,** Longitudinal resistivity for nonaligned (orange) and aligned (green) graphene/hBN devices. Solid curves: low $T$ = 10 K. Dashed: 200 K. Inset: Optical image of the SL device in **b**. **c**, $T$ dependent resistivity, Δρ, at a fixed $n$ = -1 x $10^{12}$ $cm^{-2}$ for four SL devices and the nonaligned device (orange symbols). Error bars are smaller than the data points. Inset: Device schematic and measurement scheme. The top illustration is a moiré pattern arising from 1.8 % lattice mismatch in aligned graphene (blue) and hBN (grey) crystals.

Figure 1b plots the resistivity $\rho_{xx}$ as a function of doping $n$ for two of our graphene devices at 10 K (solid lines) and 200 K (dashed). One of them is the reference, nonaligned sample (orange curves) whilst the other has a misalignment angle close to 0 and λ ≈ 15 nm (green). At low $T$, both devices exhibit comparable values of $\rho_{xx}$ for small $n$, with sharp peaks at zero doping and the resistivity that drops off rapidly with increasing $n$ for both electrons and holes (positive and negative $n$, respectively). The measured $\rho_{xx}$ are rather similar except for additional satellite peaks which occur in this SL at $n = \pm n_0 = 8/\sqrt{3}\lambda^2$ because of secondary Dirac points located at the edges of the SL Brillouin zone. At 200 K, however, the two devices exhibit remarkably different behavior even for $|n| < |n_0|$. In nonaligned graphene, the resistivity at 200 K is only marginally larger than that at 10 K. This weak $T$ dependence stems from the low electron-phonon coupling intrinsic to graphene's stiff atomic lattice. In stark contrast, the SL device exhibits a huge increase in $\rho_{xx}$, which is accompanied by a pronounced electron-hole asymmetry. Such a behavior cannot be attributed to electron scattering on thermally activated holes[7,8,9] because the effect is much stronger for doping away from the main Dirac point where the Fermi energy $\epsilon_{\rm F} > k_{\rm B}T$ and the system behaves as a metal rather than gapless semiconductor. To compare devices with different electronic quality, we analyzed the $T$ dependent part of resistivity, $\Delta\rho$, by subtracting $\rho_{xx}$ at the base $T$ of 10 K from the measured data: $\Delta\rho = \rho_{xx}(T) - \rho_{xx}$ (10 K). We have chosen 10 K to avoid an obscuring contribution from mesoscopic fluctuations at lower $T$. Fig. 1c plots $\Delta\rho(T)$ for the studied devices at a fixed density of holes. There is a huge excess resistivity in graphene SL's which grows rapidly with the moiré period. As shown below, this behavior can accurately be described by a dominant contribution from umklapp e-e scattering.

Fig. 2a details our observations by plotting $\rho_{xx}$ as a function of $n$ (normalized by $n_0$) for four SL devices, focusing on the doping level $0.2n_0 \leq |n| \leq 0.7n_0$, away from the Dirac points, miniband edges and van Hove singularities. In this range of $n$, the reconstruction of the Dirac spectrum is weak and thermal excitations of carriers with the opposite sign of effective mass can be neglected[9]. The solid and dashed lines in Fig. 2a represent $\rho_{xx}$ at 10 K and 100 K, respectively, whereas the colored shaded areas emphasize changes in resistivity. Notably, the electron-hole asymmetry increases with $n$ and, also, becomes more pronounced with increasing $\lambda$. Such asymmetry is absent in our reference graphene at any $T$. To emphasize this observation, Fig. 2b plots $\Delta\rho$(100 K) for these SL devices.

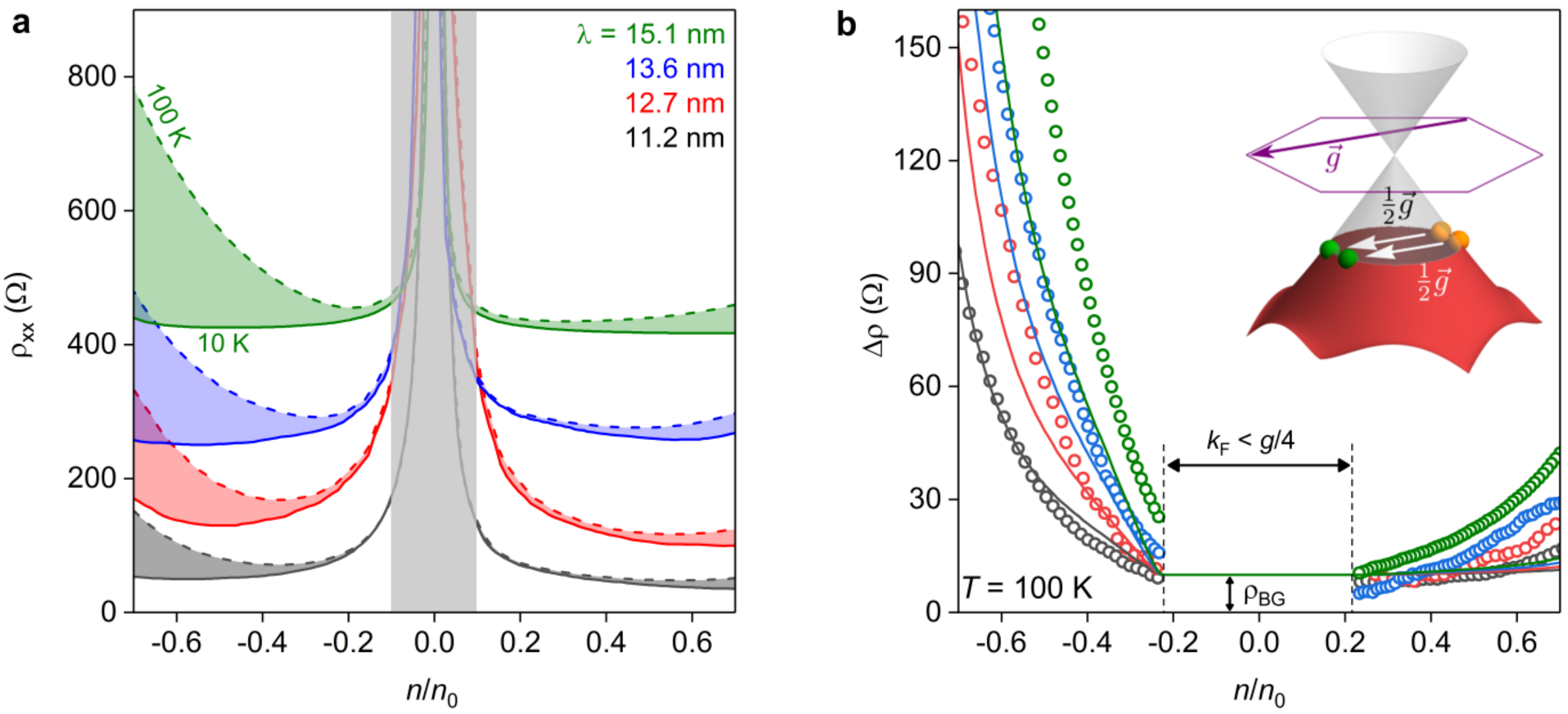

**Figure 2| Electron-electron scattering and its electron-hole asymmetry in graphene superlattices.** **a**, Resistivity for different $\lambda$ (color coded) as a function of $n$. Their $n_0$ were between 2 and 3.7 $\times10^{12}$ cm$^{-2}$. Solid curves: 10 K. Dashed: 100 K. The curves for $\lambda$ = 13.6 and 15.1 nm are offset for clarity by 200 and 400 Ohm, respectively. The color shaded areas emphasize the $T$-dependent parts of $\rho_{xx}$ for different $\lambda$. Data close to the neutrality points $n$ = 0 and $\pm$ $n_0$ are omitted (grey shading) because they exhibit activated behavior of little interest for this study. **b**, Open circles show experimental $\Delta\rho(T)$ (same color coding as in **a**). The error bars are smaller than the symbols. Solid curves: calculated Uee contribution (no fitting parameters). Note that a small density-independent offset of $\rho_{BG}$ = 10 $\Omega$ has been added to the theoretical curves to account for the resistivity generated by scattering at acoustic phonons in the Bloch–Grüneisen regime[20,21,22] . The inset depicts an umklapp

process for the threshold density $n_c$ such that $\boldsymbol{k_F} = \boldsymbol{g}/\boldsymbol{4}$ where the momentum transferred to the SL corresponds to the exact backscattering of a pair of electrons (orange and green balls).

To explain the observed behavior, we model Uee scattering for Dirac electrons in either the conduction ($s = +$) or valence ($s = -$) band of graphene using perturbation theory (PT) in both e-e Coulomb interaction and moiré SL potential. The use of the PT approach is justified by considering that, in the range of densities $0.2n_0 \leq |n| \leq 0.6n_0$, the SL inflicts only a weak change on the Dirac spectrum and the Dirac velocity[16,13] as, for example, shown previously in ARPES studies of graphene/hBN heterostructures[23,24]. In the PT approach one can envisage an Uee process as an e-e scattering event in which one of the electrons scatters into a state on the opposite side of the Fermi circle, whereas the other goes into an intermediate state with a much larger momentum (and therefore off the energy shell) and then is returned back to the Fermi line by Bragg scattering off the SL. The overall amplitude of such a process is accounted for by four Feynman diagrams, in which Bragg scattering occurs before/after an e-e collision and involves either the first or second electron,

$$M_{ss'}^{\vec{g}_m} = M_{ss'}^{\vec{g}_m}(\vec{k}_1,\vec{k}_2,\vec{k}_3,\vec{k}_4) + M_{ss'}^{-\vec{g}_m\,*}(\vec{k}_3,\vec{k}_4,\vec{k}_1,\vec{k}_2) + M_{ss'}^{\vec{g}_m}(\vec{k}_2,\vec{k}_1,\vec{k}_4,\vec{k}_3) + M_{ss'}^{-\vec{g}_m\,*}(\vec{k}_4,\vec{k}_3,\vec{k}_2,\vec{k}_1) \qquad (2)$$

$$M_{ss'}^{\vec{g}_m}(\vec{k}_1,\vec{k}_2,\vec{k}_3,\vec{k}_4) =$$

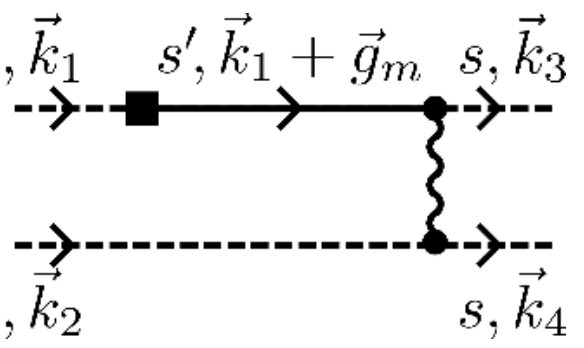

Here ∿∿∿ stands for the screened Coulomb interaction, ■ stands for the SL perturbation leading to Bragg scattering with momentum transfer $\hbar\vec{g}$, and ——— $= (\varepsilon - s'v|\vec{k}_1 + \vec{g}_m|)^{-1}$ describes the electronic propagator of the intermediate virtual state, where $v$ is the Dirac velocity in graphene and $s' = \pm$ refers to the conduction and valence band, respectively. To account for the SL scattering, we employ the previously developed model[13,14] to describe electron scattering with the shortest six moiré SL reciprocal lattice vectors,

$$\vec{g}_{m=0,\cdots 5} = (-\sin\left[\phi + \frac{\pi m}{3}\right], \cos\left[\phi + \frac{\pi m}{3}\right])g \quad \text{where} \quad g = \frac{4\pi}{\sqrt{3}\lambda} \quad \text{and} \quad \phi = \arctan\left[\frac{\sin\theta}{\delta + 1 - \cos\theta}\right].$$

Hence, for the first diagram in Eq. (2), the intermediate state has a wavevector $\vec{p}' = \vec{k}_1 + \vec{g}_m$, and the matrix element for SL scattering is

$$\blacksquare \equiv W(\vec{g}_m) = \frac{1}{2}[U_0 h_+ + i(-1)^m U_3 h_- + (-1)^m U_1 h_1], \qquad (3)$$

with $h_\pm = 1 \pm ss' e^{i\left(\vartheta_{\vec{k}_1} - \vartheta_{\vec{p}'}\right)}$ and $h_1 = s e^{\mathrm{i}\left(\vartheta_{\vec{k}_1} - \frac{\pi m}{3}\right)} + s' e^{\mathrm{i}\left(\frac{\pi m}{3} - \vartheta_{\vec{p}'}\right)}$, determined by the chirality of the electron states and the sublattice structure of the SL Hamiltonian[13,14] ($\vartheta_{\vec{k}}$ is the angle between $\vec{k}$ and the x-axis) and $U_i$ are the phenomenological parameters controlling the SL potentials. We use $U_0 = 8.5$ meV, $U_1 = -17$ meV and $U_3 = -14.7$ meV, which were determined from the previous independent study of transverse magnetic focusing in graphene/hBN superlattices[12].

For the Coulomb interaction in the first diagram of Eq. (2), we use[20,25]

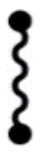

$$\equiv V(\vec{g}_m) = \frac{1 + ss' e^{i(\vartheta_{\vec{p}'} - \vartheta_{\vec{k}_3})}}{2} \frac{2\pi e^2/\kappa}{|\vec{k}_2 - \vec{k}_4| + q_{\mathrm{TF}}} \frac{1 + ss' e^{i(\vartheta_{\vec{k}_2} - \vartheta_{\vec{k}_4})}}{2}, \qquad (4)$$

with the Thomas-Fermi wavevector $q_{\mathrm{TF}} = \frac{4e^2 k_{\mathrm{F}}}{v\kappa}$, and the dielectric constant of hBN, $\kappa \approx 3.2$. Then, the first diagram in Eq. (2) is given by

$$M_{ss'}^{\vec{g}_m}(\vec{k}_1, \vec{k}_2, \vec{k}_3, \vec{k}_4) = \frac{W(\vec{g}_m) V(\vec{g}_m)}{sv|\vec{k}_1| - \tilde{s}v|\vec{p}'|}\ .$$

To determine the Uee contribution to resistivity, $\rho_{\mathrm{Uee}}$, we use the Boltzmann transport theory[26] assuming the thermal energy $k_{\mathrm{B}}T < \epsilon_{\mathrm{F}}$ (Supplementary Section 3), which yields the tensor with $\alpha, \beta = x, y$.

$$\rho_{\mathrm{Uee}}^{\alpha\beta} = \frac{h}{e^2} \frac{(k_B T)^2}{24\pi^2 v^4 k_F^2} \sum_{m=0,\cdots,5} g_m^\alpha g_m^\beta \int \left| \sum_{s'=\pm} M_{ss'}^{\vec{g}_m} \right|^2 \frac{d\vartheta_{\vec{k}_1} d\vartheta_{\vec{k}_3}}{|\sin \vartheta_{24}|} \qquad (5).$$

This expression was derived using the approximation $\vec{k}_i \approx k_F(\cos[\vartheta_{\vec{k}_i}], \sin[\vartheta_{\vec{k}_i}])$, where $k_i$ are related by Eq. (1), and the scattering angle $\vartheta_{24}$ is such that $\cos(\vartheta_{24}) = \cos(\vartheta_{\vec{k}_1} - \vartheta_{\vec{k}_3}) - \frac{g}{k_F}(\frac{g}{2k_F} + \sin\left(\vartheta_{\vec{k}_1} - \phi - \frac{\pi m}{3}\right) - \sin(\vartheta_{\vec{k}_3} - \phi - \frac{\pi m}{3}))$. Because of the $C_3$ rotational symmetry of graphene superlattices the resistivity tensor is isotropic, that is, $\rho_{\mathrm{Uee}}^{\alpha\beta} = \rho_{\mathrm{Uee}} \delta^{\alpha\beta}$. We note that Uee scattering can occur only above the threshold $k_{\mathrm{F}} > g/4$ (inset of Fig. 2b) which arises from the fact that all scattering states must be in the vicinity of the Fermi level, $k_i \approx k_F$ in Eq. (1) and yields the critical density $n_c = n_0 \pi / 8\sqrt{3} \approx 0.227 n_0$ below which umklapp e-e scattering is not allowed.

Using the SL parameters $U_i$ stated above, we calculated $\rho_{\mathrm{Uee}}$ for the specific experimental parameters in Fig. 2a. The results (solid curves in Fig. 2b) are in good agreement with the experiment, which is particularly impressive considering that no fitting parameters were used. Note that the deviations between the experiment and theory for electron doping in Fig. 2b are mostly due to a limited accuracy of our analytical method as the full numerical analysis shows (see Fig. S5a). Furthermore, the analytical theory, Eq. (5), suggests that close to the threshold density $\rho_{\mathrm{Uee}} \propto (|n| -$ $n_c)^{3/2}$, which stems from the interplay between the size of the phase space available for Uee scattering and the 'chirality factor' [e.g. $(1+e^{i(\vartheta_{\vec{k}_2}-\vartheta_{\vec{k}_4})})/2$ in Eq. (4)] which suppresses the amplitude of backscattering[27]. The large asymmetry between $\rho_{Uee}$ for electrons and holes arises from the fact that kinematic constraints dictate that the electron Bragg scattering by the SL must be almost backscattered (inset of Fig 2b). The probability for such backscattering,

$$P \sim |U_1 - sU_3|^2, \tag{6}$$

is much higher in the valence band ($s = -1$) than the conduction band ($s = 1$) for the given $U_1$ and $U_3$ used in Eq. (3) so that the Uee process is much more effective for hole rather than electron doping. Note that this feature of Uee distinguishes itself from other scattering mechanisms including the potential disorder in the moiré SL[28], which results in almost electron-hole symmetric $\rho_{xx}$ within the density range $-0.7n_0 < n < 0.7n_0$.

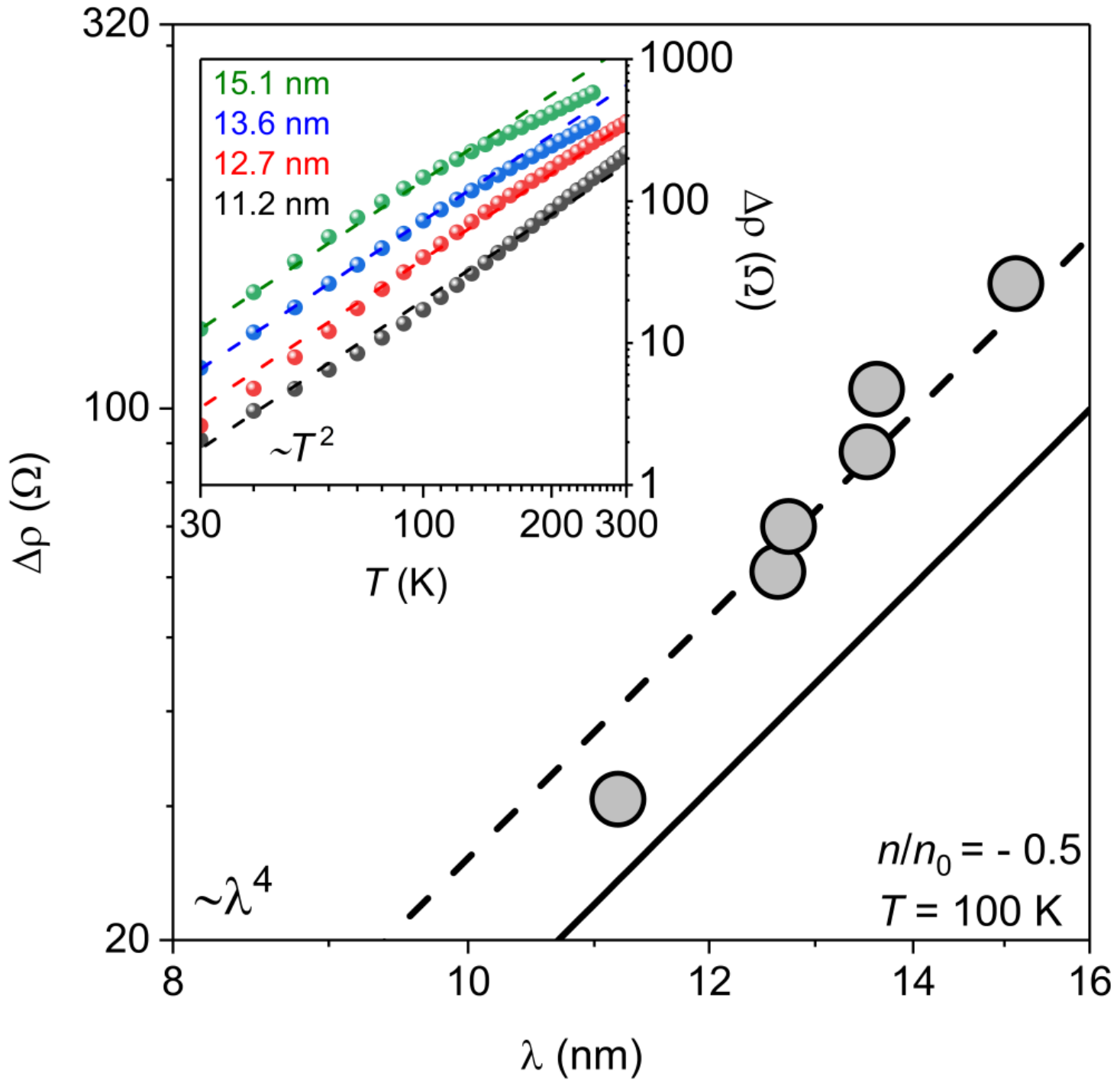

**Figure 3| Characteristics of umklapp electron-electron scattering.** *T* dependent part of resistivity as a function of moiré period for $n = -0.5\ n_0$ in all our six SL devices. The circles are experiment data; the dashed line is the best fit of a $\lambda^4$ dependence to the data; and the solid line is the calculated Uee contribution to the resistivity (also $\propto \lambda^4$). Inset: Symbols are experimental data for the SL devices (color-coded). The dashed lines are $T^2$ fits to the experimental data. Logarithmic scales are used in both the main plot and the inset. Standard deviations in our measurements are smaller than all the symbols.

Equation (5) predicts two further signatures of Uee scattering. First, for a given $|n| > n_c$, umklapp processes should result in a strong dependence on $g$ such that $\rho_{Uee} \propto \lambda^4$ for a fixed $n/n_0$, where the scaling behavior is determined entirely by the dependence of the Uee matrix element $M_{ss'}^{\vec{g}_m}$ on the superlattice period. Fig. 3 shows that this dependence describes well the observed behavior of $\Delta\rho$. This unusually strong dependence is one of the reasons why nonaligned devices with small SL periods do not exhibit any discernable umklapp resistivity. Second, Eq. (5) yields a quadratic dependence typical for electron-electron scattering in the Fermi liquid theory, $\rho_{\mathrm{Uee}} \propto T^2$, in agreement with the experimental behavior plotted in the inset of Fig. 3. The $T^2$ behavior holds over a wide $T$ range for all our SL devices showing the dominance of Uee scattering. However, at high $T >$ 150 K, one can see some deviations from the $T^2$ dependence. We attribute those to the thermal excitation of carriers with the opposite sign of effective mass, resulting in deviations of resistivity from the values described by Eq. (5). Indeed, these deviations become stronger as we approach either the main Dirac point ($|n| < 0.3n_0$) or van Hove singularities[13] ($|n| > 0.6n_0$) where scattering at thermally excited carriers of the opposite polarity starts playing a role.

Finally, let us mention that we analyzed the normal and umklapp (due to the moiré SL) scattering of electrons at acoustic phonons in graphene. The normal electron-phonon scattering, studied in detail previously[20,21,22], can result in approximately a 10 Ω contribution for the relevant $n$ at 100 K (the value used as an offset in Fig. 2b) and up to 30 Ω at 300 K. An additional scattering on hBN's phonons may also contribute to the deviations. This is discussed in Supplementary Sections 3C-E, where we consider a possibility that electrons scatter off acoustic phonons in graphene and hBN by transferring additional momentum $\hbar\vec{g}$ to the moiré SL (Fig. S5c). When analyzing such processes, we took into account the intrinsic electron-phonon coupling (deformation potential) in graphene, piezoelectric coupling with deformations in hBN and dynamical variations of the moiré potential due to a mutual displacement of graphene and hBN, which are caused by vibrations of the two crystals. We find that the calculated phonon-induced umklapp resistivity is much smaller and has different $T$- and $\lambda$-dependences, as compared to those caused by Uee scattering and observed experimentally (Fig. S5).

To conclude, umklapp e-e scattering in graphene superlattices determines their resistivity over a wide range of temperatures and carrier densities. This dominance of umklapp scattering is unique to graphene/hBN superlattices due to the small size of their Brillouin zone and exceptionally weak electron-phonon coupling. The umklapp scattering is particularly important for hole doping. We expect that umklapp e-e scattering should strongly influence electron transport in twisted graphene bilayers where superlattice effects have also been predicted[29,30] and recently observed[31,32,33,34,35].

**REFERENCES**

1. Bass, J., Pratt, W. P. & Schroeder, P. A. The temperature-dependent electrical resistivities of the alkali metals. *Rev. Mod. Phys.* **62,** 645–744 (1990).

2. Gasparov, V. A. & Huguenin, R. Electron-phonon, electron-electron and electron-surface scattering in metals from ballistic effects. *Adv. Phys.* **42,** 393–521 (1993).

3. Messica, A. *et al.* Suppression of Conductance in Surface Superlattices by Temperature and Electric Field. *Phys. Rev. Lett.* **78,** 705–708 (1997).

4. Overend, N. *et al.* Giant magnetoresistance and possible miniband effects in periodic magnetic fields. *Phys. B Condens. Matter* **249–251,** 326–329 (1998).

5. Kato, M., Endo, A., Katsumoto, S. & Iye, Y. Two-dimensional electron gas under a spatially modulated magnetic field: A test ground for electron-electron scattering in a controlled environment. *Phys. Rev. B* **58,** 4876–4881 (1998).

6. Kato, M., Endo, A., Sakairi, M., Katsumoto, S. & Iye, Y. Electron-electron umklapp process in two-dimensional electron gas under a spatially alternating magnetic field. *J. Phys. Soc. Japan* **68,** 1492–1495 (1999).

7. Kashuba, A. B. Conductivity of defectless graphene. *Phys. Rev. B* **78,** 85415 (2008).

8. Fritz, L., Schmalian, J., Müller, M. & Sachdev, S. Quantum critical transport in clean graphene. *Phys. Rev. B* **78,** 85416 (2008).

9. Nam, Y., Dong-Keun, K., Soler-Delgado, D. & Morpurgo, A. F. Electron–hole collision limited transport in charge-neutral bilayer graphene. *Nat. Phys.* **13,** 1207-1214 (2017).

10. Rice, T. M., Robinson, N. J. & Tsvelik, A. M. Umklapp scattering as the origin of *T*-linear resistivity in the normal state of high-$T_c$ cuprate superconductors. *Phys. Rev. B* **96,** 220502 (2017).

11. Aleiner, I. L. & Agam, O. Saturation of strong electron–electron umklapp scattering at high temperature. *Ann. Phys.* **385,** 716–728 (2017).

12. Lee, M. *et al.* Ballistic miniband conduction in a graphene superlattice. *Science* **353,** 1526-1529 (2016).

13. Wallbank, J. R., Patel, A. A., Mucha-Kruczyński, M., Geim, A. K. & Fal'ko, V. I. Generic miniband structure of graphene on a hexagonal substrate. *Phys. Rev. B* **87,** 245408 (2013).

14. Wallbank, J. R., Mucha-Kruczyński, M., Chen, X. & Fal'ko, V. I. Moiré superlattice effects in

graphene/boron-nitride van der Waals heterostructures. *Ann. Phys.* **527,** 359–376 (2015).

15. Yankowitz, M. *et al.* Emergence of superlattice Dirac points in graphene on hexagonal boron nitride. *Nat. Phys.* **8,** 382–386 (2012).

16. Ponomarenko, L. A. *et al.* Cloning of Dirac fermions in graphene superlattices. *Nature* **497,** 594–597 (2013).

17. Dean, C. R. *et al.* Hofstadter's butterfly and the fractal quantum Hall effect in moire superlattices. *Nature* **497,** 598–602 (2013).

18. Hunt, B. *et al.* Massive Dirac fermions and Hofstadter butterfly in a van der Waals heterostructure. *Science* **340,** 1427-1430 (2013).

19. Krishna Kumar, R. *et al.* High-temperature quantum oscillations caused by recurring Bloch states in graphene superlattices. *Science* **357,** 181-184 (2017).

20. Hwang, E. H. & Das Sarma, S. Acoustic phonon scattering limited carrier mobility in two-dimensional extrinsic graphene. *Phys. Rev. B* **77,** 115449 (2008).

21. Chen, J-H., Jang, C., Xiao, S., Ishigami, M. & Fuhrer, M. S. Intrinsic and extrinsic performance limits of graphene devices on SiO2. *Nat. Nano* **3,** 206–209 (2008).

22. Wang, L. *et al.* One-dimensional electrical contact to a two-dimensional material. *Science* **342,** 614-617 (2013).

23. Wang, E. *et al.* Gaps induced by inversion symmetry breaking and second-generation Dirac cones in graphene/hexagonal boron nitride. *Nat. Phys.* **12,** 1111–1115 (2016).

24. Wang, E. *et al*. Electronic structure of transferred graphene/h-BN van der Waals heterostructures with nonzero stacking angles by nano-ARPES. *J. Phys. Condens. Matter* **28,** 444002 (2016).

25. Hwang, E. H. & Das Sarma, S. Dielectric function, screening, and plasmons in two-dimensional graphene. *Phys. Rev. B* **75,** 205418 (2007).

26. Ziman, J. M. Electrons And Phonons: The theory of transport phenomena in solids - Sec 9.14. Oxford University Press (1960).

27. Castro Neto, A. H., Guinea, F., Peres, N. M. R., Novoselov, K. S. & Geim, A. K. The electronic properties of graphene. *Rev. Mod. Phys.* **81,** 109–162 (2009).

28. DaSilva, A. M., Jung, J., Adam, S. & MacDonald, A. H. Transport and particle-hole asymmetry in graphene on boron nitride. *Phys. Rev. B* **91,** 245422 (2015).

29. Lopes dos Santos, J. M. B., Peres, N. M. R. & Castro Neto, A. H. Graphene bilayer with a twist: Electronic structure. *Phys. Rev. Lett.* **99,** 256802 (2007).

30. Bistritzer, R. & MacDonald, A. H. Transport between twisted graphene layers. *Phys. Rev. B* **81,** 245412 (2010).

31. Li, G. *et al.* Observation of van Hove singularities in twisted graphene layers. *Nat. Phys.* **6,** 109-113 (2010).

32. Kim, K. *et al.* Tunable moiré bands and strong correlations in small-twist-angle bilayer graphene. *Proc. Natl. Acad. Sci.* **114,** 3364-3369 (2017).

33. Sanchez-Yamagishi, J. D. *et al.* Quantum Hall effect, screening, and layer-polarized insulating states in twisted bilayer graphene. *Phys. Rev. Lett.* **108,** 76601 (2012).

34. Rode, J. C., Smirnov, D., Schmidt, H. & Haug, R. J. Berry phase transition in twisted bilayer graphene. *2D Mater.* **3,** 35005 (2016).

35. Cao, Y. *et al.* Unconventional superconductivity in magic-angle graphene superlattices. *Nature* **556,** 43-50 (2018).

# Supplementary Information for: Excess resistivity in graphene superlattices caused by umklapp electron-electron scattering

J. R. Wallbank,[1] R. Krishna Kumar,[1,2] M. Howill,[1,2] Z. Wang,[2] G. H. Auton,[1] J. Birkbeck,[1,2] A. Mishchenko,[1,2] L. A. Ponomarenko,[3] K. Watanabe,[4] T. Taniguchi,[4] K. S. Novoselov,[1,2] I. L. Aleiner,[5] A. K. Geim,[1,2] and V. I. Fal'ko[1,2]

[1]*National Graphene Institute, University of Manchester, Manchester M13 9PL, UK*
[2]*School of Physics & Astronomy, University of Manchester, Oxford Road, Manchester M13 9PL, UK*
[3]*Department of Physics, University of Lancaster, Lancaster LA1 4YW, United Kingdom*
[4]*National Institute for Materials Science, 1-1 Namiki, Tsukuba 305-0044, Japan*
[5]*Physics Department, Columbia University, New York, NY 10027, USA*

## S1. DEVICE FABRICATION

The graphene/hexagonal boron nitride (hBN) devices presented in the main text were fabricated following the methods reported previously [1]. First, we used a dry-transfer method [2] for assembling the heterostructures. We obtained monolayer graphene and few layer hBN by mechanical exfoliation of graphite and bulk hexagonal boron-nitride crystals on to a silicon/silicon dioxide (Si/SiO2) wafer. After the appropriate flakes were identified, we used a polymer membrane attached to the tip of a micromanipulator to assemble the heterostructure. This was performed on a rotating stage which allowed us to control the relative angle between two crystal lattices to a precision of about $0.5°$. We first assembled monolayer graphene on the hBN substrate. During this step, we used the rotating stage to try to align their crystallographic axes in order to produce a moiré superlattice [3]. Because the flakes cleave preferentially along their crystallographic directions, the straight edges of the few layer crystals tell us their relative orientation. However, alignment of straight edges does not guarantee alignment of the crystal axes because of the two types of edges that exist (arm-chair or zig-zag). Therefore, we performed atomic force microscopy (AFM) measurements of the graphene/hBN stack to check for an underlying moiré superlattice [4]. If a superlattice was obtained, we then placed a second hBN flake on top of the stack to in order to fully encapsulate the graphene flake and preserve its intrinsic electronic quality [1]. We then used standard methods in electron beam lithography to fabricate the hall bar geometry and define quasi-one dimensional contacts to the graphene edge [5, 6]. The structural properties of the moiré superlattice were also confirmed in transport experiments by measuring the position of secondary Dirac points [3] and frequency of Brown-Zak oscillations [7].

## S2. CHARACTERISING SUPERLATTICE DEVICES

The superlattice devices presented in the main text have a moiré period ranging between 10-15 nm. The resulting period ($\lambda$) depends on the misalignment angle ($\theta$) between hBN and graphenes' crystallographic axes. During fabrication, we try to align straight edges of the exfoliated flakes to the highest accuracy possible. However, the crystals tend to rotate slightly during the transfer procedure which makes it difficult to control the resulting alignment angle $\theta < 0.5°$. The superlattices produced in this way could have periods varying between 10-15 nm. Fortunately, the exact period can be determined directly from transport experiments. Figure 1a plots resistivity ($\rho_{xx}$) as a function of gate voltage ($V_G$) for one of our superlattice devices at 100,K (Fig. S1a inset). Aside from the sharp peak in $\rho_{xx}$ around $V_G = 0\,\mathrm{V}$, there are two additional peaks occurring at around $\pm 50\,\mathrm{V}$. These satellite peaks signify secondary Dirac points that are generic to graphene/hBN moiré superlattices [8]. To determine $\lambda$ directly, we perform magnetotransport experiments and study Brown-Zak oscillations [7]. These high-temperature oscillations are periodic in $1/B$ with a frequency ($B_F$) that depends only on the size of the superlattice unit cell ($S$) and corresponds to the condition when one flux quantum pierces it, $B_F = \phi_0/S$ (Fig. S1b inset). Therefore, by measuring their frequency we can determine $\lambda$ directly. Figure. 2 plots $\rho_{xx}$ as a function of $B$ measured for fixed $VG = 30\,\mathrm{V}$ at $100\,\mathrm{K}$. The data shows strong oscillations which are periodic in $1/B$. For this particular device, $B_F$ was found to be $30.2\,\mathrm{T}$ which corresponds to $\lambda = 12.6\,\mathrm{nm}$. Alternatively, $\lambda$ can be determined by measuring the carrier density where secondary Dirac points occur [9–11] ($n_0$) since they correspond to a doping level of four electrons per superlattice unit cell $n_0 = 4/S$ (see main text). We cross checked the $\lambda$ obtained by both methods and found agreement to within $\sim 1\%$.

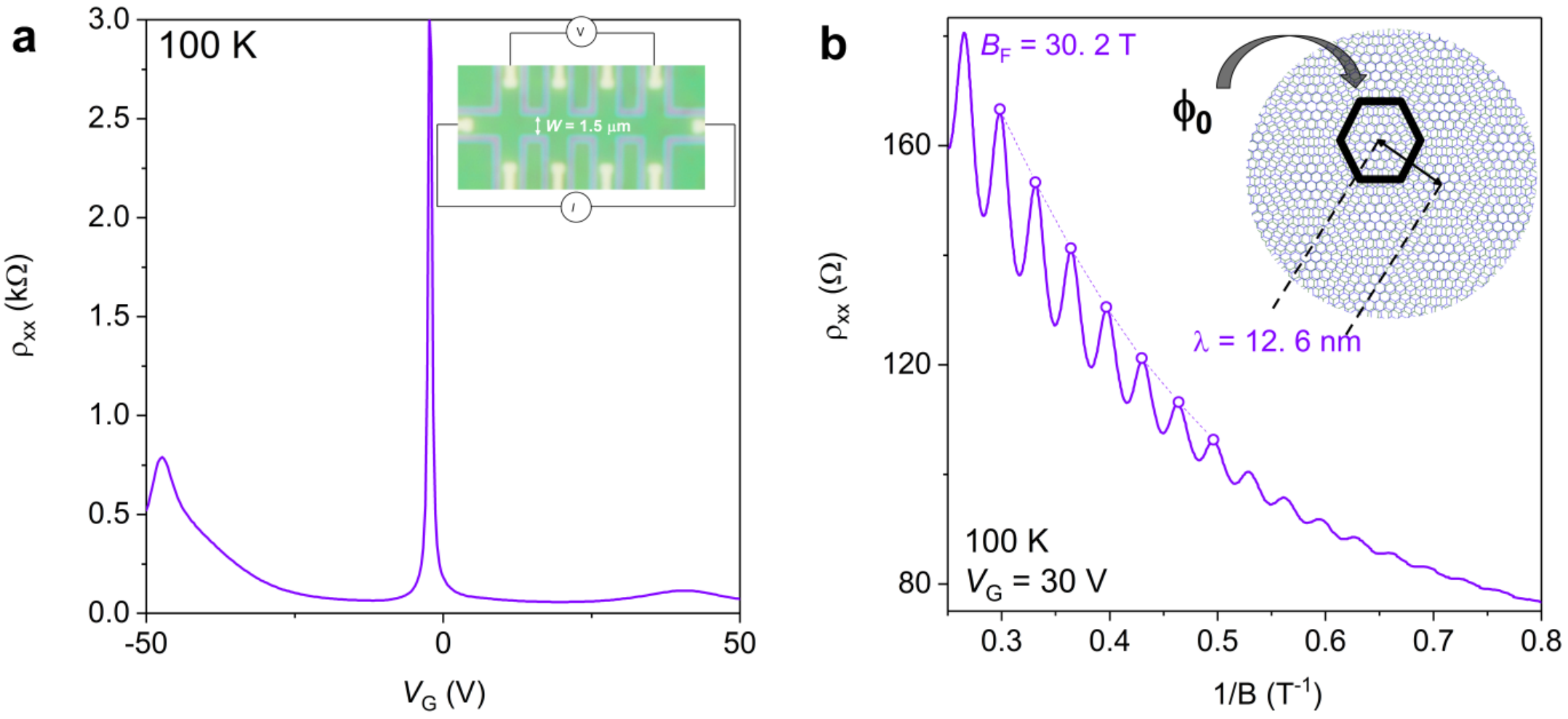

FIG. S1: **Measurement of moiré period by studying Brown-Zak oscillations.** (**a**), $\rho_{xx}(V_G)$ at 100 K for one of our superlattice devices. Inset; optical image and measurement schematic of the corresponding device. (**b**), Magnetoresistance $\rho_{xx}(1/B)$ for $V_G = 30$ V. Solid lines are experimental data whilst the open circles are hand drawn data points to extract the fundamental frequency $B_F$. Inset; An illustration of the graphene/hBN moiré superlattice (outlined by the black hexagon). The oscillations have a fundamental frequency corresponding to one flux quantum piercing the moiré unit cell.

## S3. UMKLAPP SCATTERING MECHANISMS IN GRAPHENE/HBN HETEROSTRUCTURES

In this section we describe the Boltzmann transport calculation for the resistivity generated by Uee scattering, and compare it to the calculated contributions produced by either a SL coupling to acoustic phonons at the interface between the graphene and hBN crystals, or, the scattering of electron at piezoelectric potentials generated by acoustic phonons in the hBN. We will show that the phonon contributions to the resistivity are more than an order of magnitude smaller than that produced by Uee scattering, and are therefore neglected in the main text. We will use $\hbar = k_B = 1$ throughout.

### A. Hamiltonian for graphene/hBN superlattices

First we describe the SL Hamiltonian for graphene/hBN and the consequent reconstruction of graphene's Dirac cone into a series of minibands. To model graphene/hBN we employ the SL Hamiltonian [8, 12],

$$H = v\boldsymbol{k}\cdot\boldsymbol{\sigma} + \sum_{m=0,\cdots 5}\left[U_0 + (-1)^m\left(iU_3\sigma_3 + U_1\frac{\boldsymbol{G}_m\times\hat{\boldsymbol{z}}}{G}\cdot\boldsymbol{\sigma}\right)\right]e^{i\boldsymbol{g}_m\cdot\boldsymbol{r}} \tag{S.1}$$

Here the first term describes Dirac electrons in graphene's $K$-valley ($\sigma_{i=0,1,2,3}$ and $\boldsymbol{\sigma} = (\sigma_1, \sigma_2)$ are Pauli matrices), while the second term is the SL potential [8, 12], with strengths $U_{i=0,1,3}$ given in the main text. Also, $\boldsymbol{G}_{m=0,\cdots 5}$ are the shortest six graphene reciprocal lattice vectors obtained by in-plane $m\pi/3$ rotation of $\boldsymbol{G}_0 = (0,1,0)G$ where $G = \frac{4\pi}{\sqrt{3}a}$, $a = 2.46$ Å, and $\boldsymbol{g}_m$ are the moiré reciprocal lattice vectors defined in the main text.

The numerically calculated miniband structure of Hamiltonian (S.1) is displayed in Fig. S2 (also see insets of Fig. 1 and Fig. 2 main text). For energies $|\epsilon| \lesssim vg/2$ (densities $|n| \lesssim n_0$), indicated by the black arrow in Fig. S2, the bandstructure is only weakly affected by the SL potentials. A quantitative estimate for the effect of the SL potentials on the low-energy bandstructure can be obtained using second order perturbation theory. Expanding to second order in wavevector $k$ also, we obtain the following effective Hamiltonian for the $|\epsilon| \lesssim vg/2$ dispersion of graphene/hBN in the K-valley,

$$H_{\text{eff}} = v\boldsymbol{k}\cdot\boldsymbol{\sigma} + w_0\sigma_0 + w_1\frac{\boldsymbol{k}}{g}\cdot\boldsymbol{\sigma} + w_2\frac{k^2}{g^2}\sigma_0, \tag{S.2}$$

$$w_0 = \frac{12\delta U_1 U_3}{vg\sqrt{\delta^2+\theta^2}}, \quad w_1 = -\frac{6}{vg}(U_0^2 + U_3^2) - \frac{12\delta^2 U_1^2}{vg(\delta^2+\theta^2)}, \quad w_2 = \frac{24\delta U_1 U_3}{vg\sqrt{\delta^2+\theta^2}}. \tag{S.3}$$

In the main text we neglected the SL reconstruction of the bandstructure (only the first term in Eq. (S.3) was retained). The remaining terms are all small in $U_i^2/(vg)$. For example, using the SL potentials $U_i$ given in the main text, $\theta = 0$, and $\delta = 1.8\%$, the SL correction to the Dirac velocity is $w_1/g = -0.3\,\text{eV\AA}$ compared to $v = 6.6\,\text{eV\AA}$ for plain graphene. Nevertheless, SL reconstruction of the bandstructure can become more significant near the edges of the density range $|n| \lesssim 0.7n_0$ studied in the main text. The effect of this on the calculated Uee contribution to the resistivity is studied in more detail in supplementary section S3 B 2.

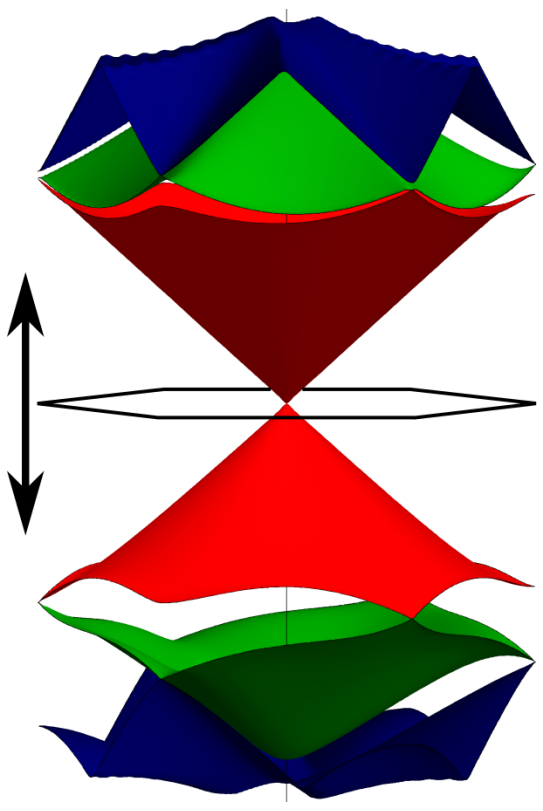

FIG. S2: The minibandstructure shown with in the SL Brillouin zone (black rhombus), calculated for $\theta = 0$, $\delta = 1.8\%$ and $U_0 = 8.5\,\text{meV}$, $U_1 = -17\,\text{meV}$, and $U_3 = -14.7\,\text{meV}$ [13]. The black double arrow indicates the approximate range of energies for which the bandstructure remains Dirac-like.

### B. Boltzmann transport calculation for resistivity produced by umklapp electron-electron scattering

#### *1. Leading-order term*

Here we describe the calculation of the Uee contribution to the resistivity, $\rho_{\text{Uee}}$, leading to Eq. 5 in the main text. We neglect the reconstruction of the Dirac spectrum and wavefunction by the SL, which corresponds to a leading order perturbation theory calculation in the small parameter $U_i/(vg)$. A numerical calculation to include beyond leading order affects is described in section S3 B 2. Throughout we assume $T \ll \epsilon_F$.

All of our calculations will employ the linearised Boltzmann equation [14, 15] for the electron distribution function,

$$e\boldsymbol{E}\cdot\boldsymbol{v}(\boldsymbol{k}_1)\frac{\partial f_0(\epsilon_{\boldsymbol{k}_1})}{\partial \epsilon_{\boldsymbol{k}_1}} = I\{\psi_{\boldsymbol{k}_1}\}, \tag{S.4}$$

where, $\boldsymbol{E} = (E, 0)$ is the applied electric field which is assumed (without loss of generality) to point in the x-direction, $\boldsymbol{v}(\boldsymbol{k}_1) = sv(\cos(\vartheta_{\boldsymbol{k}_1}), \sin(\vartheta_{\boldsymbol{k}_1}))$ is the electron velocity with $s = +/-$ for the conduction/valence band, and the r.h.s of Eq. (S.4) is the collision integral (described below). The electron distribution is expanded as,

$$f(\boldsymbol{k}) = f_0(\epsilon_{\boldsymbol{k}}) - \frac{f_0(\epsilon_{\boldsymbol{k}})}{d\epsilon_{\boldsymbol{k}}}\psi_{\boldsymbol{k}}, \tag{S.5}$$

where

$$f_0(\epsilon) = \frac{1}{e^{(\epsilon-\epsilon_F)/T} + 1}. \tag{S.6}$$

is the equilibrium electron distribution and the unknown function $\psi_{\boldsymbol{k}}$ varies slowly in energy.

The resistivity, $\rho$, is then obtained using,

$$\begin{aligned}\rho^{-1} &= \frac{4esv}{E}\int \frac{d\boldsymbol{k}}{(2\pi)^2}\cos(\vartheta_{\boldsymbol{k}})f(\boldsymbol{k}) \\ &\approx \frac{se}{\pi^2 E}\int d\vartheta_{\boldsymbol{k}} k_F \cos(\vartheta_{\boldsymbol{k}})\psi_{\boldsymbol{k}},\end{aligned} \tag{S.7}$$

where the factor 4 in the first line accounts for the spin-valley degeneracy. In the second line the integration is performed at the Fermi-level, and we have used Eq. (S.5) together with the fact that $f_0(\epsilon_{\boldsymbol{k}})/d\epsilon_{\boldsymbol{k}}$ can be approximated using a Dirac delta-function at the Fermi-level when $T \ll \epsilon_F$.

For Uee scattering, the collision integral scattering in Eq. (S.4) reads [14, 15],

$$I\{\psi_{\boldsymbol{k}_1}\} = \frac{4\times 2\pi}{T}\sum_m \int \frac{d\boldsymbol{k}_2 d\boldsymbol{k}_3 d\boldsymbol{k}_4}{(2\pi)^6} |\sum_{s'} M_{ss'}^{\boldsymbol{g}_m}|^2 \delta(\Delta\epsilon)(2\pi)^2\delta(\Delta\boldsymbol{k})\frac{1}{16}\prod_{i=1,\cdots,4}\operatorname{sech}\left(\frac{\epsilon_i-\epsilon_F}{2T}\right)\times\{\psi_{\boldsymbol{k}_4}+\psi_{\boldsymbol{k}_3}-\psi_{\boldsymbol{k}_2}-\psi_{\boldsymbol{k}_1}\},$$

$$\Delta\epsilon = sv|\boldsymbol{k}_1| + sv|\boldsymbol{k}_2| - sv|\boldsymbol{k}_3| - sv|\boldsymbol{k}_4|, \qquad \Delta\boldsymbol{k} = \boldsymbol{k}_1+\boldsymbol{k}_2-\boldsymbol{k}_3-\boldsymbol{k}_4+\boldsymbol{g}_m \tag{S.8}$$

To obtain the electron distribution function from Eqs. (S.4) and (S.8), we use the ansatz,

$$\psi_{\boldsymbol{k}} = \alpha k^x, \tag{S.9}$$

with $\alpha$ an unknown parameter, and $\boldsymbol{k} = (k^x, k^y)$. An equation for $\alpha$ is obtained by multiplying both sides of Eq. (S.4) by $k_1^x$ and integrating to yield,

$$\alpha^{-1} \approx \sum_m \frac{g_m^{x\,2} k_F^2}{128\pi^4 eE} I_k I_\vartheta \tag{S.10}$$

where the integral has been split into the radial and angular parts,

$$I_k = \frac{1}{T}\int dk_1 dk_2 dk_3 dk_4 \prod_{i=1,\cdots,4}\operatorname{sech}\left(\frac{vk_i-\epsilon_F}{2T}\right)\delta(\Delta\epsilon) = \frac{32\pi^2T^2}{3v^4},$$

$$I_\vartheta = \int d\vartheta_{\boldsymbol{k}_1} d\vartheta_{\boldsymbol{k}_2} d\vartheta_{\boldsymbol{k}_3} d\vartheta_{\boldsymbol{k}_4} |\sum_{s'} M_{ss'}^{\boldsymbol{g}_m}|^2 \delta(\Delta\boldsymbol{k}),$$

and the approximation $\boldsymbol{k}_i \approx k_F(\cos(\vartheta_{\boldsymbol{k}_i}), \sin(\vartheta_{\boldsymbol{k}_i}))$ is used in $I_\vartheta$. To perform a partial integration [15] of $I_\vartheta$, we change variables from $\vartheta_{\boldsymbol{k}_2}$, $\vartheta_{\boldsymbol{k}_4}$, to $(\boldsymbol{k}_4 - \boldsymbol{k}_2)$,

$$d\vartheta_{\boldsymbol{k}_2} d\vartheta_{\boldsymbol{k}_4} = \frac{d(\boldsymbol{k}_4-\boldsymbol{k}_2)}{k_F^2 \sin\vartheta_{24}}, \qquad \cos(\vartheta_{24}) = \cos(\vartheta_{\boldsymbol{k}_1}-\vartheta_{\boldsymbol{k}_3}) - \frac{g}{k_F}\left(\frac{g}{2k_F} + \sin(\vartheta_{\boldsymbol{k}_1}-\phi-\pi m/3) - \sin(\vartheta_{\boldsymbol{k}_3}-\phi-\pi m/3)\right), \tag{S.11}$$

with $\phi = \arctan(\frac{\sin(\theta)}{\delta+1-\cos(\theta)})$ the angle between the principal directions of moiré SL and the graphene lattice. This allows us to use the momentum-conserving Dirac-delta function in $I_\vartheta$ to obtain,

$$I_\vartheta = \int \frac{|M|^2 d\vartheta_{\boldsymbol{k}_1} d\vartheta_{\boldsymbol{k}_3}}{k_F^2|\sin\vartheta_{24}|}. \tag{S.12}$$

Note that in integral (S.12) $\vartheta_{\boldsymbol{k}_2}$ and $\vartheta_{\boldsymbol{k}_4}$ (entering $M_{ss'}^{\boldsymbol{g}_m}$) are chosen to solve $\Delta\boldsymbol{k} = 0$ (there will typically be two solutions to this equation, and the values of the integrand for each must be summed). Then $\rho_{\rm Uee}$, given in Eq. 5 of the main text, is obtained by using Eq. (S.9) in Eq. (S.7) with $\alpha$ obtained from (S.10).

### 2. *Beyond leading-order terms*

Effects beyond the leading order in $U_i/(vg)$ consist of (i) the SL reconstruction of graphene's bandstructure, and, (ii) the SL reconstruction of graphene's wavefunction. Here we describe how these effects can be included in the calculation of $\rho_{\rm Uee}$. Later (Fig. S5) we will show that they do not produce a significant effect on the resistivity. This stems from the smallness of $U_i/(vg)$ and also a partial cancellation of the effect of the SL reconstruction of graphene's bandstructure (the reduction in Fermi velocity tends to increase the calculated resistivity) and the SL reconstruction of graphene's wavefunctions (which tends to reduce the matrix element for scattering).

For densities with in the first minibands we can account for the SL reconstruction of graphene's bandstructure by replacing the Dirac velocity $v$ in $\rho_{\rm Uee}$ (Eq. (5) main text) with a numerically calculated average value at the Fermi level, defined from the density of states (DoS) using,

$$<v>_F = \frac{2\sqrt{|n|}}{\sqrt{\pi}{\rm DoS}}. \tag{S.13}$$

The numerically calculated DoS and the ratio $<v>_F /v$ are presented in Fig. S3.

The reconstruction of the graphene's wavefunction can be fully taken into account by replacing the lowest order perturbation theory expression for the Uee scattering matrix element, Eq. (2) in the main text, by an expression based on the numerically calculated eigenvectors of Hamiltonian (S.1). That is, in Eq. (5) in the main text we replace

$$\sum_{s'} M_{ss'}^{\boldsymbol{g}_m} \to \sum_{s_1,s_2,s_3,s_4} \sum_{\boldsymbol{G}_1,\boldsymbol{G}_2,\boldsymbol{G}_3,\boldsymbol{G}_4} \delta_{\boldsymbol{G}_1+\boldsymbol{G}_2-\boldsymbol{G}_3-\boldsymbol{G}_4,\boldsymbol{g}_m} \frac{1+s_1 s_3 e^{i(\theta_{\boldsymbol{k}_1+\boldsymbol{G}_1}-\theta_{\boldsymbol{k}_3+\boldsymbol{G}_3})}}{2} \frac{2\pi e^2/\kappa}{|\boldsymbol{k}_1+\boldsymbol{G}_1-\boldsymbol{k}_3-\boldsymbol{G}_3|+q_{\mathrm{TF}}}$$
$$\times \frac{1+s_2 s_4 e^{i(\theta_{\boldsymbol{k}_2+\boldsymbol{G}_2}-\theta_{\boldsymbol{k}_4+\boldsymbol{G}_4})}}{2} A_{s_1,\boldsymbol{G}_1}(\boldsymbol{k}_1) A_{s_2,\boldsymbol{G}_2}(\boldsymbol{k}_2) A^*_{s_3,\boldsymbol{G}_3}(\boldsymbol{k}_3) A^*_{s_4,\boldsymbol{G}_4}(\boldsymbol{k}_4). \quad \text{(S.14)}$$

Here $\boldsymbol{G}_i$ are Bragg vectors of the SL (summed over all $|\boldsymbol{G}_i| < G_{\max}$ with large enough $G_{\max}$ for convergence) and $s_i = \pm$ are band indexes. The $A_{s_i,\boldsymbol{G}_i}(\boldsymbol{k}_i)$ are the coefficients of the eigenfunction, $\psi_{s,\boldsymbol{k}_i}$, of Hamiltonian (S.1) expanded in basis of graphene plain-waves (and are obtained from the eigenvectors of Hamiltonian (S.1) computed in the same basis),

$$\psi_{s,\boldsymbol{k}_i}(\boldsymbol{r}) = \sum_{s_i,\boldsymbol{G}_i} A_{s_i,\boldsymbol{G}_i}(\boldsymbol{k}_i) \frac{1}{\sqrt{2}} \begin{pmatrix} 1 \\ s_i e^{i\theta_{\boldsymbol{k}_i+\boldsymbol{G}_i}} \end{pmatrix} e^{i(\boldsymbol{k}_i+\boldsymbol{G}_i)\cdot\boldsymbol{r}}. \quad \text{(S.15)}$$

We note that the expression for $\sum_{s'} M_{ss'}^{\boldsymbol{g}_m}$ presented in the main text is recovered from Eq. (S.14) if first order perturbation theory expressions are used for the coefficients $A_{s_i,\boldsymbol{G}_i}(\boldsymbol{k}_i)$.

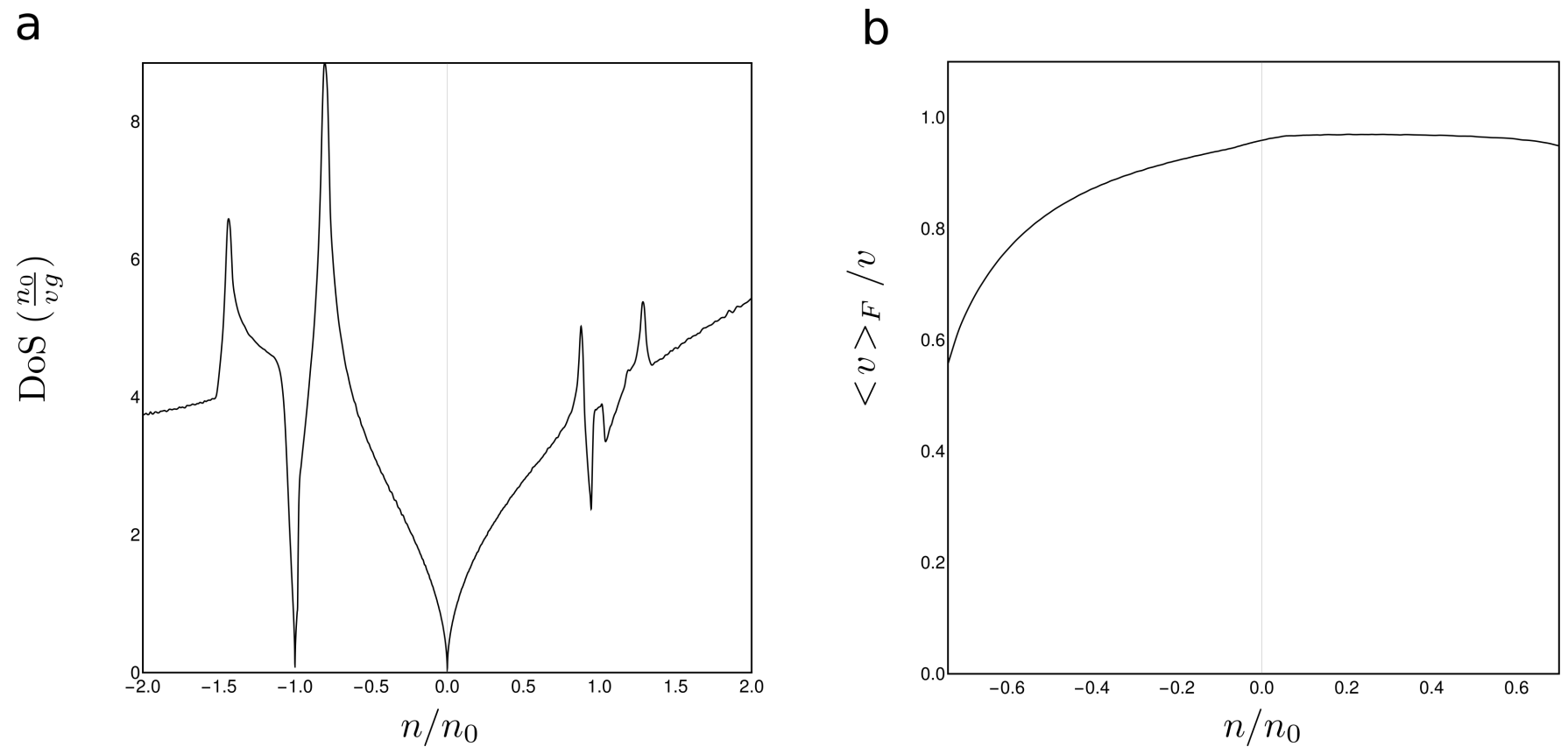

FIG. S3: **(a)** The DoS numerically calculated from the bandstructure in Fig. S2, and **(b)**, the ratio of the associated average Fermi velocity to the Dirac velocity of plain graphene $<v>_F /v$.

### C. Umklapp scattering with acoustic phonon modes in graphene and hBN

It has been noticed that straining either graphene or hBN unilaterally immediately leads to an anisotropic appearance of the moiré SL, as well as dislocations in the two crystals inflicting dislocation like deformations of the moiré pattern [16]. Figure S4 (a) illustrates how a small local deformation mimicking a linear combination of LA and TA-type sound waves in graphene manifests itself in the deformations in the moiré pattern. The resulting deformation of the SL potential produces an umklapp electron-phonon scattering in which the interaction of graphene's Dirac electrons with the SL allows them to emit in-plane acoustic phonons in both the graphene and hBN layers with wavevectors supplemented by the SL reciprocal lattice vectors. At the same time, the out-of-plane acoustic modes can alter the separation between graphene and hBN affecting the overall strength of the moiré SL potentials, additionally providing a channel for electron umklapp scattering on out-of-plane phonon modes also. Here we present a calculation of the resistivity generated by such umklapp scattering on both in-plane, $\rho_{Ue-ph}$, and out-of-plane, $\rho_{Ue-zph}$, acoustic phonons.

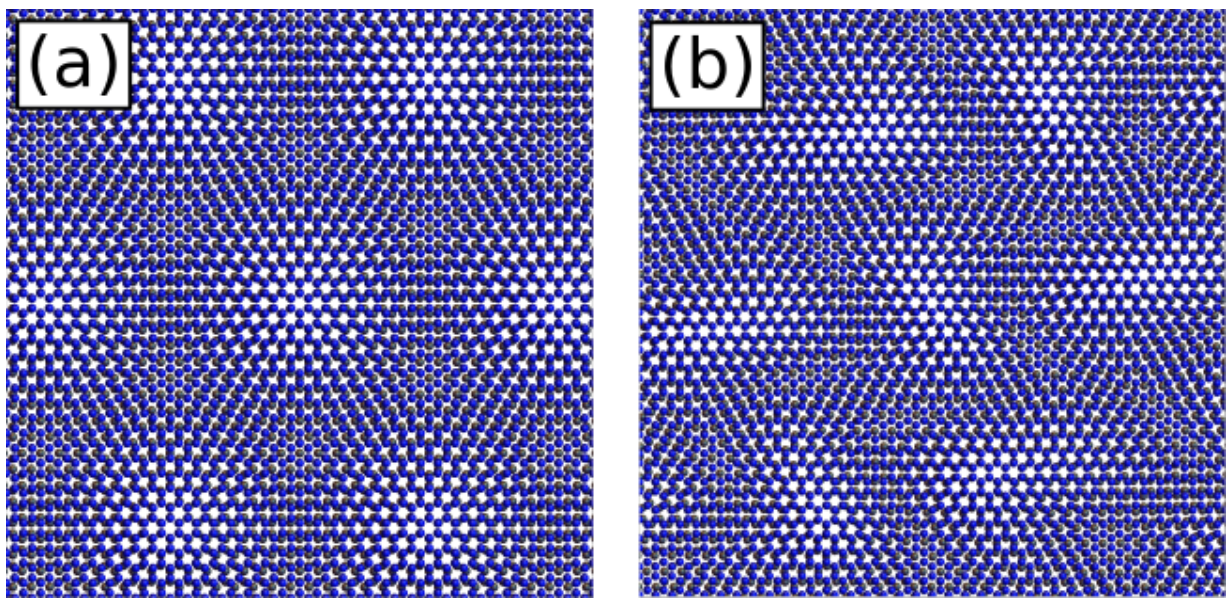

FIG. S4: **Umklapp scattering on in-plane acoustic phonons** The moiré pattern of pristine graphene on hBN **(a)**, or with deformations **(b)** mimicking the acoustic phonons produced at finite temperatures ($\boldsymbol{u}_{\rm Gr} = 0.2a(\boldsymbol{q}_1/q_1)\sin(\boldsymbol{q}_1\cdot\boldsymbol{r}) + 0.2a(\hat{\boldsymbol{z}}\times\boldsymbol{q}_2/q_2)\sin(\boldsymbol{q}_2\cdot\boldsymbol{r})$, $\boldsymbol{q}_1 = (0.4, 1)g$, $\boldsymbol{q}_2 = (0.2, -0.2)g$).

### 1. *Hamiltonian for electron-phonon SL coupling*

To describe the electron-phonon coupling, we generalise the SL potential in Hamiltonian (S.1) to account for deformations [12, 16] in the graphene and hBN lattices,

$$H = v\boldsymbol{k}\cdot\boldsymbol{\sigma} + \sum_{m=0,\cdots 5}\left[U_0 + (-1)^m\left(iU_3\sigma_3 + U_1\frac{\boldsymbol{G}_m\times\hat{\boldsymbol{z}}}{G}\cdot\boldsymbol{\sigma}\right)\right]e^{i\boldsymbol{g}_m\cdot\boldsymbol{r}}e^{-i\boldsymbol{G}_m\cdot\boldsymbol{u}(\boldsymbol{r})},$$
$$U_i(u_z) = U_i(0) - u_z(\boldsymbol{r})\partial_z U_i(z). \tag{S.16}$$

Here $\boldsymbol{u}(\boldsymbol{r}) = \boldsymbol{u}_{\rm G} - \boldsymbol{u}_{\rm hBN}$ is a small local deformation of the two crystals ($\boldsymbol{u}_{\rm G}$ and $\boldsymbol{u}_{\rm hBN}$), where the in-plane component of $\boldsymbol{u}(\boldsymbol{r})$ affect the phase of the SL potential, and the out-of-plane component, $u_z$, affects the size of the parameters $U_i$ controlling the strength of the SL potential. Also, we estimate,

$$\partial_z U_i = \eta G U_i \tag{S.17}$$

with $\eta \sim 1$.

We obtain the Hamiltonian for the SL electron-phonon coupling by retaining terms in Hamiltonian (S.16) at first order in the displacements and then quantising using the phonon field operators,

$$H_{\text{e-ph}} = -i\sum_{\boldsymbol{q},m}\left[U_0 + (-1)^m\left(iU_3\sigma_3 + U_1\frac{\boldsymbol{G}_m\times\hat{\boldsymbol{z}}}{G}\cdot\boldsymbol{\sigma}\right)\right]\left\{\sum_\nu\sqrt{\frac{1}{2\rho_m\omega^\nu_{\boldsymbol{q}}}}\boldsymbol{G}_m\cdot\hat{1}_\nu(\boldsymbol{q})\hat{A}_{\boldsymbol{q},\nu} + \sqrt{\frac{1}{2\rho_m\omega_\perp(q)}}G\hat{A}_q\right\}\frac{e^{i(\boldsymbol{q}+\boldsymbol{g}_m)\cdot\boldsymbol{r}}}{L}. \tag{S.18}$$

Here $L^2$ is the area of the flake. Within the curly braces, the first term is the coupling with the in-plane phonon modes, with $\nu$ running over LA and TA modes in both graphene and hBN, $\hat{A}_{\boldsymbol{q},\nu} = (b_{\boldsymbol{q},\nu} + b^\dagger_{-\boldsymbol{q},\nu})$ is written in terms of the phonon creation/annihilation operators ($b^\dagger_{\boldsymbol{q},\nu}/b_{\boldsymbol{q},\nu}$), $\rho_m = 7.6\times10^{-7}$kg m$^2$ is the mass density, and the phonon polarisations are $\hat{1}_\nu(\boldsymbol{q}) = \pm\boldsymbol{q}/q$ or $\hat{1}_\nu(\boldsymbol{q}) = \pm\hat{\boldsymbol{z}}\times\boldsymbol{q}/q$ for LA or TA mode phonons (and the $+/-$ is used for G/hBN). For the hBN phonon energies we use $\omega^\nu_{\boldsymbol{q}} = s_\nu q$ with $s_{\rm LA} = 0.12\,$eVÅ or $s_{\rm TA} = 0.08\,$eVÅ [17], while for graphene we use $\omega^\nu_{\boldsymbol{q}} = c_\nu q$ with $c_{\rm LA} = 0.15\,$eVÅ or $c_{\rm TA} = 0.09\,$eVÅ [18]. The second term within the braces describes the coupling with the out-of-plane phonon mode in which the graphene and hBN are out of phase with each other. Here we use $\omega_\perp(q) \approx \Delta$ with $\Delta = 0.01\,$eV [19] for the phonon dispersion in the momentum range of interest ($q \lesssim g$). Note that we have implicitly assumed the top hBN layer to be mechanically de-coupled from the rest of the hBN slab. This results in softer phonon modes, and hence higher resistivities $\rho_{Ue-ph}$ and $\rho_{Ue-zph}$, than would otherwise be the case.

### 2. *Resistivity produced by umklapp scattering with in-plane acoustic phonons*

To calculate the resistivity generated by umklapp scattering with in-plane acoustic phonons, we use Boltzmann transport equation (S.4) with the collision integral [14],

$$I\{\psi_{\boldsymbol{k}}\} = \sum_{\eta,m,\nu}\int\frac{d\boldsymbol{k}'}{(2\pi)^2}\mathcal{W}(\boldsymbol{g}_m)\frac{dN}{d\omega^\nu_{\boldsymbol{q}}}[f_0(\epsilon_{\boldsymbol{k}'}) - f_0(\epsilon_{\boldsymbol{k}})](\psi_{\boldsymbol{k}'} - \psi_{\boldsymbol{k}})\eta\delta(\epsilon_{\boldsymbol{k}} - \epsilon_{\boldsymbol{k}'} + \eta\omega^\nu_{\boldsymbol{q}}), \tag{S.19}$$

$$\boldsymbol{q} = \boldsymbol{k} - \boldsymbol{k}' + \boldsymbol{g}_m \tag{S.20}$$

Here $\nu$ sums over LA and TA phonon modes in both graphene and hBN, $m = 0, \cdots 5$, $\eta = \pm$ accounts for phonon emission and absorption, and $N = (e^{\omega^{\nu}_{\boldsymbol{q}}/T} - 1)^{-1}$ is the equilibrium phonon distribution. The intrinsic scattering probability, $\mathcal{W}(\boldsymbol{g}_m)$, for a Dirac electron, $|s\boldsymbol{k}\rangle$, to scatter with an in-plane acoustic phonon $|\nu\boldsymbol{q}\rangle$, is calculated using the first term with in the braces of Hamiltonian (S.18),

$$\begin{aligned}\mathcal{W}(\boldsymbol{g}_m) =& 2\pi L^2 |\langle s\boldsymbol{k}', \nu\boldsymbol{q}|H_{\text{e-ph}}|s\boldsymbol{k}\rangle|^2. \\ =& \frac{\pi\left(\boldsymbol{g}_m \cdot \hat{1}^{\boldsymbol{q}}_{\nu}\right)^2}{\rho_m \omega^{\nu}_{q}} \left|U_0 h_+ + i(-1)^m U_3 h_- + (-1)^m s U_1 h_1\right|^2\end{aligned} \tag{S.21}$$

where $h_{\pm} = \frac{1 \pm e^{i(\vartheta_{\boldsymbol{k}} - \vartheta_{\boldsymbol{k}'})}}{2}$ and $h_1 = \frac{e^{i(\vartheta_{\boldsymbol{k}} - \frac{\pi m}{3})} + e^{-i(\vartheta_{\boldsymbol{k}'} - \frac{\pi m}{3})}}{2}$ as per the main text, and we note that the electron occupancy factors in Eq. (S.19) only allow intra-band scattering ($T \ll \epsilon_F$).

To simplify collision integral (S.19), we use the energy conserving Dirac-delta function to write,

$$f_0(\epsilon_{\boldsymbol{k}'}) = f_0(\epsilon_{\boldsymbol{k}} + \eta\omega^{\nu}_{\boldsymbol{q}}) \approx f_0(\epsilon_{\boldsymbol{k}}) + \eta\omega^{\nu}_{\boldsymbol{q}} \frac{df_0(\epsilon_{\boldsymbol{k}})}{d\epsilon_{\boldsymbol{k}}}, \tag{S.22}$$

where the higher order terms in the Taylor series are neglect due to the smoothness of the integrand. Then integrating both sides of the Boltzmann transport equation (S.4) over $|\boldsymbol{k}|$ we obtain,

$$eEsv\cos(\vartheta) = \frac{k_F}{2\pi^2 v} \sum_{\nu,m} \int d\vartheta' \mathcal{W}(\boldsymbol{g}_m) \frac{dN}{d\omega^{\nu}_{\boldsymbol{q}}} \omega^{\nu}_{\boldsymbol{q}} (\psi_{\boldsymbol{k}'} - \psi_{\boldsymbol{k}}). \tag{S.23}$$

Next, we expand $\psi_{\boldsymbol{k}}$ in terms of its angular harmonics,

$$\psi_{\boldsymbol{k}} = \sum_n \hat{\psi}^n_k e^{in\vartheta_{\boldsymbol{k}}}, \tag{S.24}$$

so that equation (S.23) become,

$$1 = \frac{k_F}{2\pi^3 seEv^2} \sum_{\nu,m,n} \int d\vartheta d\varphi \mathcal{W}(\boldsymbol{g}_m) \frac{dN}{d\omega^{\nu}_q} \omega^{\nu}_{\boldsymbol{q}} (e^{in\varphi} - 1) e^{i(n-1)\vartheta_{\boldsymbol{k}}} \hat{\psi}^n_{k_F}, \tag{S.25}$$

where we have define the scattering angle $\varphi = \vartheta_{\boldsymbol{k}'} - \vartheta_{\boldsymbol{k}}$.

We will now proceed by making different approximations in the two distinct regimes of (i) $k_F \ll g$, and, (ii) $0 < k_F < g/2$.

**(i) For $k_F \ll g$:**
In this limit we use the fact that $\omega^{\nu}_q \approx \omega^{\nu}_g$. Then the only $m$ dependence in Eq. (S.25) is contained in $\mathcal{W}(\boldsymbol{g}_m)$, which can be summed explicitly using,

$$\sum_m \mathcal{W}(\boldsymbol{g}_m) = \frac{3\pi\left(\boldsymbol{G}_0 \cdot \hat{1}^{\boldsymbol{g}_0}_{\nu}\right)^2}{\rho_m \omega^q_{\nu}} \left(U_0^2 + U_3^2 + U_1^2 + (U_0^2 - U_3^2)\cos(\varphi)\right). \tag{S.26}$$

Performing the integration and summing on $n$ in Eq. (S.25) yields,

$$\frac{E}{\hat{\psi}^1_{k_F}} = -\frac{6k_F}{ev^2 M} W_0^2 \sum_{\nu} |\boldsymbol{G}_0 \cdot \hat{1}^{\boldsymbol{g}_0}_{\nu}|^2 \frac{dN}{d\omega^{\nu}_b}, \tag{S.27}$$

so that using Eq. (S.7) and also

$$\frac{dN}{d\omega^{\nu}_{\boldsymbol{q}}} = \frac{-1}{2T(\cosh\left(\frac{\omega^{\nu}_{\boldsymbol{q}}}{T}\right) - 1)}, \text{ and } \left|\boldsymbol{G}_0 \cdot \hat{1}^{\boldsymbol{g}_0}_{\nu}\right|^2 \approx \frac{G^2}{\delta^2 + \theta^2} \times \begin{cases}\delta^2, & \text{for LA phonons} \\ \theta^2, & \text{for TA phonons}\end{cases}, \tag{S.28}$$

we obtain the resistivity

$$\rho_{\text{Ue-ph}} = \frac{h}{e^2} \frac{3G^2 \mathcal{W}_0^2}{4v^2 \rho_m (\delta^2 + \theta^2)} \sum_{\nu} \frac{\delta^2 \delta_{\nu,\text{LA}} + \theta^2 \delta_{\nu,\text{TA}}}{T\cosh(\frac{\omega^b_{\nu}}{T}) - T}, \qquad \mathcal{W}_0^2 = \frac{1}{2}U_0^2 + \frac{3}{2}U_3^2 + U_1^2. \tag{S.29}$$

**(ii) For $0 < k_F < b/2$:**
To proceed in the general case, $k_F \lesssim g/2$, we make use of the following two symmetries of the scattering probability (explicitly including the angular dependencies in the arguments of $\mathcal{W}$),

$$\mathcal{W}(\boldsymbol{g}_m, \vartheta_{\boldsymbol{k}}, \vartheta_{\boldsymbol{k}'}) = \mathcal{W}(\boldsymbol{g}_{m+2}, \vartheta_{\boldsymbol{k}} + 2\pi/3, \vartheta_{\boldsymbol{k}'} + 2\pi/3), \quad \text{and} \quad \mathcal{W}(\boldsymbol{g}_0, \vartheta_{\boldsymbol{k}}, \vartheta_{\boldsymbol{k}'}) = \mathcal{W}(\boldsymbol{g}_3, -\vartheta_{\boldsymbol{k}}, -\vartheta_{\boldsymbol{k}'}). \tag{S.30}$$

Using these the sum over $m$ in Eq. (S.25) can be evaluated to yield,

$$\frac{E}{\hat{\psi}^1_{k_F}} = \frac{3k_F}{\pi^3 sev^2} \sum_\nu \int d\varphi d\vartheta_{\boldsymbol{k}} \mathcal{W}(\boldsymbol{g}_0) \frac{dN}{d\omega^\nu_q} \omega^\nu_q (\cos(\varphi) - 1), \hat{\psi}^1_{k_F} \tag{S.31}$$

with $\boldsymbol{q} = \boldsymbol{k} - \boldsymbol{k}' - \boldsymbol{g}_0$. In principle the integral in Eq. (S.31) could be preformed numerically, however we prefer to produce an analytical answer using the following two simplifying assumptions: (i) that $T > \omega^\nu_g$ so that $dN/d\omega^\nu_{\boldsymbol{q}} \approx -T/(\omega^\nu_{\boldsymbol{q}})^2$, and, (ii) we set $c_{\rm TA} = c_{\rm LA}$ and $s_{\rm TA} = s_{\rm LA}$. Then Eq. (S.31) is reduced to,

$$\begin{aligned} \frac{E}{\hat{\psi}^1_{k_F}} &= -\int d\varphi d\vartheta_{\boldsymbol{k}} \frac{3k_F T G^2}{ev^2\pi^2\rho_m} \left(\frac{1}{c^2_{\rm LA}} + \frac{1}{s^2_{\rm LA}}\right) |U_0 h_+ + iU_3 h_- + sU_1 h_1|^2 \frac{(\cos(\varphi) - 1)}{|\boldsymbol{k} - \boldsymbol{k}' + \boldsymbol{g}_0|^2} \\ &= -\frac{6k_F G^2 T}{sev^2\rho_m b^2} \left(\frac{1}{c^2_{\rm LA}} + \frac{1}{s^2_{\rm LA}}\right) \mathcal{W}^2_{k_F}, \end{aligned}$$

where

$$\mathcal{W}^2_{k_F} = \frac{2U_0^2}{(x+1)^2} + U_1^2 + \frac{2(x+2)}{x(x+1)^2}\left(U_3 - \frac{s\delta U_1\sqrt{1-x^2}}{\sqrt{\delta^2+\theta^2}}\right)^2, \quad x = \sqrt{1 - (2k_F/g)^2}.$$

In the first line we used $(\boldsymbol{G}_m \cdot \hat{1}^{\boldsymbol{q}}_{\rm LA})^2 + (\boldsymbol{G}_m \cdot \hat{1}^{\boldsymbol{q}}_{\rm TA})^2 = g^2$ to perform the sum over the phonon modes, while in the second line the integrals are performed by making the substitutions $z = e^{i\varphi}$, $w = e^{i\vartheta_{\boldsymbol{k}}}$ and using the residue theorem.
Then the resistivity is obtained using Eq. (S.7),

$$\rho_{\text{Ue-ph}} \approx \frac{h}{e^2} \frac{3G^2 T}{2v^2\rho_m} \frac{\mathcal{W}^2_{k_F}}{g^2} \left(\frac{1}{c^2_{\rm LA}} + \frac{1}{s^2_{\rm LA}}\right). \tag{S.32}$$

### 3. *Scattering off out-of-plane phonon modes*

It is easy to adapt Eq. (S.29) describing the resistivity generated by the SL coupling to the in-plane phonon modes in the $k_F \ll g$ limit to the out-of-plane modes,

$$\rho_{\text{Ue-zph}} = \frac{h}{e^2} \frac{3\eta^2 G^2 W_0^2}{4v^2\rho_m} \frac{1}{T\cosh(\frac{\Delta}{T}) - T}. \tag{S.33}$$

The qualitative difference between the in-plane and out-of-plane phonons is that the out-of-plane modes are approximately non-dispersive for the range of wavevector of interest ($q \lesssim g$). This results in an insensitivity of the resistivity to the Fermi-wavevector of the heterostructure, and hence Eq. (S.33) applies for the any $k_F < g/2$ in either band (in contrast to the resistivity generated by the in-plane modes which is highly sensitive to the level of doping, Eq. (S.32)).

## D. Electron scattering off piezo-electrically coupled hBN phonons

Deformations created by acoustic phonons in a stack of hBN produce piezoelectric potentials that can scatter electrons in the graphene layer. Here we present calculations of the resistivity produced by this novel piezoelectric coupling for both normal, $\rho_{\text{e-ph}_{\text{hBN}}}$, and umklapp, $\rho_{\text{Ue-ph}_{\text{hBN}}}$, scattering events. We show that both processes only lead to small additions to the total resistivity of the device.

### 1. Piezoelectric electron-phonon coupling

We consider a single graphene layer sat on top of an $N_l$-layer hBN slab. The electrical polarisation generated by the $n^{\text{th}}$-hBN layer in the slab is given by [20–22],

$$\boldsymbol{P}_n = (-1)^n \gamma \mathcal{A} \times \hat{z}, \qquad \mathcal{A} = \begin{pmatrix} \partial_x u_n^x - \partial_y u_n^y \\ -\partial_x u_n^y - \partial_y u_n^x \end{pmatrix} \tag{S.34}$$

where $\gamma = 3.71 \times 10^{-10}$C/m [20, 21] is the $\gamma_{yyy}$ component of the piezoelectric tensor, $\boldsymbol{u}_n = (u_n^x, u_n^y)$ is by the phonon displacement on layer $n$, and the $(-1)^n$ factor accounts for the fact that the orientation of the single layer hBN layers alternate for each layer in the slab.

To obtain the potential, $\psi(\boldsymbol{r})$, generated at the graphene layer we solve Poisson's equation for the polarisation generated by hBN acoustic phonons,

$$\kappa \nabla^2 \psi = 4\pi \sum_{n=1}^{N} \nabla \cdot \boldsymbol{P}_n, \tag{S.35}$$

with $\kappa \approx 3.2$ for hBN.

$$\psi(\boldsymbol{r}, z) = \sum_{\nu, q_z, \boldsymbol{q}, n} \frac{2\pi\gamma}{\kappa} \left( \frac{1}{2N_l \rho_m \omega^{\nu}_{q,q_z}} \right)^{\frac{1}{2}} \boldsymbol{q} \cdot \hat{\zeta}^{\nu}_{\boldsymbol{q}} (-1)^n e^{i(\boldsymbol{q}\cdot\boldsymbol{r} + q_z c n)} e^{-qcn} A_{\boldsymbol{q}}, \qquad \hat{\zeta}^{\nu}_{\boldsymbol{q}} = \frac{1}{q^2} \begin{cases} \begin{pmatrix} 2q_x q_y \\ q_x^2 - q_y^2 \end{pmatrix}, & \text{for LA phonons} \\ \begin{pmatrix} q_x^2 - q_y^2 \\ -2q_x q_y \end{pmatrix}, & \text{for TA phonons} \end{cases}, \tag{S.36}$$

where $\nu$ sums over the LA and TA modes delocalised over the hBN slab, $\hat{A}_{\boldsymbol{q},\nu} = (\hat{b}_{\boldsymbol{q},\nu} + \hat{b}^{\dagger}_{-\boldsymbol{q},\nu})$, the z-direction phonon wavevector is $q_z = 2\pi j/(cN_l)$ with $j = 1, \cdots N_l$, and the hBN interlayer spacing $c = 3.4$ Å. For the phonon dispersion we use $\omega^{\nu}_{q,q_z} = \sqrt{(s_\nu q)^2 + (\Delta_0 \sin(q_z c))^2}$ with $\Delta_0 = 6.2$ meV [23, 24].

### 2. Resistivity generated by normal scattering events

First, we calculate the normal (non-umklapp) resistivity, $\rho_{\text{e-ph}_{\text{hBN}}}$, associated with e-ph coupling (S.36). This contribution to the resistivity is expected in all graphene/hBN devices regardless of the alignment between graphene and hBN. To model this contribution to the resistivity we use the Boltzmann equation S.4 with the collision integral,

$$I\{\psi_{\boldsymbol{k}}\} = \sum_{\eta,\nu,q_z} \int \frac{d\boldsymbol{k}'}{(2\pi)^2} \mathcal{W} \frac{dN}{d\omega^{\nu}_{q,q_z}} [f_0(\epsilon_{\boldsymbol{k}'}) - f_0(\epsilon_{\boldsymbol{k}})](\psi_{\boldsymbol{k}'} - \psi_{\boldsymbol{k}}) \eta \delta(\epsilon_{\boldsymbol{k}} - \epsilon_{\boldsymbol{k}'} + \eta \omega^{\nu}_{q,q_z}), \tag{S.37}$$

$$\mathcal{W} = 2\pi L^2 |\langle s\boldsymbol{k}', \nu\boldsymbol{q}, q_z | e\psi | s\boldsymbol{k} \rangle|^2, \qquad \boldsymbol{q} = \boldsymbol{k} - \boldsymbol{k}',$$

where, using (S.36), we evaluate,

$$\mathcal{W} = \frac{2\pi^3 e^2 \gamma^2 (\boldsymbol{q} \cdot \hat{\zeta}^{\nu}_{\boldsymbol{q}})^2}{\rho_M \omega^{\nu}_{q,q_z} \kappa^2 N_l} \left| \sum_n (-1)^n e^{incq_z} e^{-cnq} \right|^2 (1 + \cos(\varphi)). \tag{S.38}$$

where $\varphi = \theta_{\boldsymbol{k}} - \theta_{\boldsymbol{k}'}$. Following a similar set of steps as section S3 B, the resistivity for $vk_F \ll T$ is,

$$\rho_{\text{e-ph}_{\text{hBN}}} = \frac{h}{e^2} \sum_{\nu,q_z} \int d\varphi \frac{\pi q^2 e^2 \gamma^2}{8N_l \rho_M v^2 \kappa^2} \left| \sum_n (-1)^n e^{incq_z} e^{-cnq} \right|^2 \frac{1 - \cos^2(\varphi)}{T \cosh(\frac{\omega^{\nu}_{q,q_z}}{T}) - T}, \tag{S.39}$$

where $q = k_F \sqrt{2} \sqrt{1 - \cos(\varphi)}$.

### 3. Resistivity generated by umklapp scattering events

Here we calculate the resistivity, $\rho_{\text{Ue-ph}_{\text{hBN}}}$ generated by umklapp scattering with piezo-electrically coupled acoustic phonons in the hBN slab. The collision integral is,

$$I\{\psi_{\boldsymbol{k}}\} = \sum_{\eta,\nu,q_z,m} \int \frac{d\boldsymbol{k}'}{(2\pi)^2} \mathscr{W}(\boldsymbol{g}_m) \frac{dN}{d\omega^\nu_{\boldsymbol{q}}} [f_0(\epsilon_{\boldsymbol{k}'}) - f_0(\epsilon_{\boldsymbol{k}})](\psi_{\boldsymbol{k}'} - \psi_{\boldsymbol{k}})\eta\delta(\epsilon_{\boldsymbol{k}} - \epsilon_{\boldsymbol{k}'} + \eta\omega^\nu_{\boldsymbol{q}}), \tag{S.40}$$

$$\mathscr{W}(\boldsymbol{g}_m) = \frac{2\pi}{\hbar} L^2 |\mathscr{M}(\boldsymbol{g}_m)|^2, \qquad \boldsymbol{q} = \boldsymbol{k} - \boldsymbol{k}' + \boldsymbol{g}_m.$$

where both the piezo-electrically coupled phonon scattering and scattering on the SL are treated using perturbation theory described by the diagrams,

$$\mathscr{M}(\boldsymbol{g}_m) = [\text{diagram: } \boldsymbol{k} \text{ - - -}\blacksquare\text{—}\bullet\text{- - - } \boldsymbol{k}'] + [\text{diagram: } \boldsymbol{k} \text{ - - -}\bullet\text{—}\blacksquare\text{- - - } \boldsymbol{k}']$$

$$= \sum_{s',n} \frac{\pi\gamma\boldsymbol{q}\cdot\hat{\zeta}^\nu_{\boldsymbol{q}}(-1)^n e^{icnq_z - qcn}}{\kappa\sqrt{2N_l\rho_M\omega^\nu_{q,q_z}}} \left( \frac{(1+ss'e^{i(\vartheta_{\boldsymbol{k}+\boldsymbol{g}_m} - \vartheta_{\boldsymbol{k}'})})W(\boldsymbol{k},\boldsymbol{g}_m)}{svk - s'v|\boldsymbol{k}+\boldsymbol{g}_m|} + \frac{(1+ss'e^{i(\vartheta_{\boldsymbol{k}+\boldsymbol{g}_m} - \vartheta_{\boldsymbol{k}'})})W(\boldsymbol{k}',-\boldsymbol{g}_m)^*}{svk' - s'v|\boldsymbol{k}'-\boldsymbol{g}_m|} \right)$$

Similar to the main text, ■ represents scattering off the SL and ——— $= (\epsilon - svp)^{-1}$ represents the electronic propagator, while ▮▮▮▮▮▮▮ represents scattering off the piezoelectric potentials (Eq. (S.36)). Similar to the main text,

$$\blacksquare = W(\boldsymbol{k},\boldsymbol{g}_m) = \frac{1}{2}\left[U_0 h_+ + i(-1)^m U_3 h_- + (-1)^m U_1 h_1\right], \quad h_\pm = 1 \pm ss'e^{i(\vartheta_{\boldsymbol{k}} - \vartheta_{\boldsymbol{k}+\boldsymbol{g}_m})}, \quad h_1 = se^{i(\vartheta_{\boldsymbol{k}} - \frac{\pi m}{3})} + s'e^{i(\frac{\pi m}{3} - \vartheta_{\boldsymbol{k}+\boldsymbol{g}_m})}.$$

Following a similar set of steps as section S3 B, the resistivity for $\epsilon_F \ll T$ is,

$$\rho_{\text{Ue-ph}_{\text{hBN}}} = \frac{h}{e^2}\frac{3}{8\pi^3 v^2} \sum_{\nu,q_z} \int d\vartheta d\varphi \mathscr{W}(\boldsymbol{g}_0) \frac{\omega^\nu_{q,q_z}(1-\cos(\varphi))}{T\cosh(\frac{\omega^\nu_{q,q_z}}{T}) - T} \tag{S.41}$$

## E. Discussion

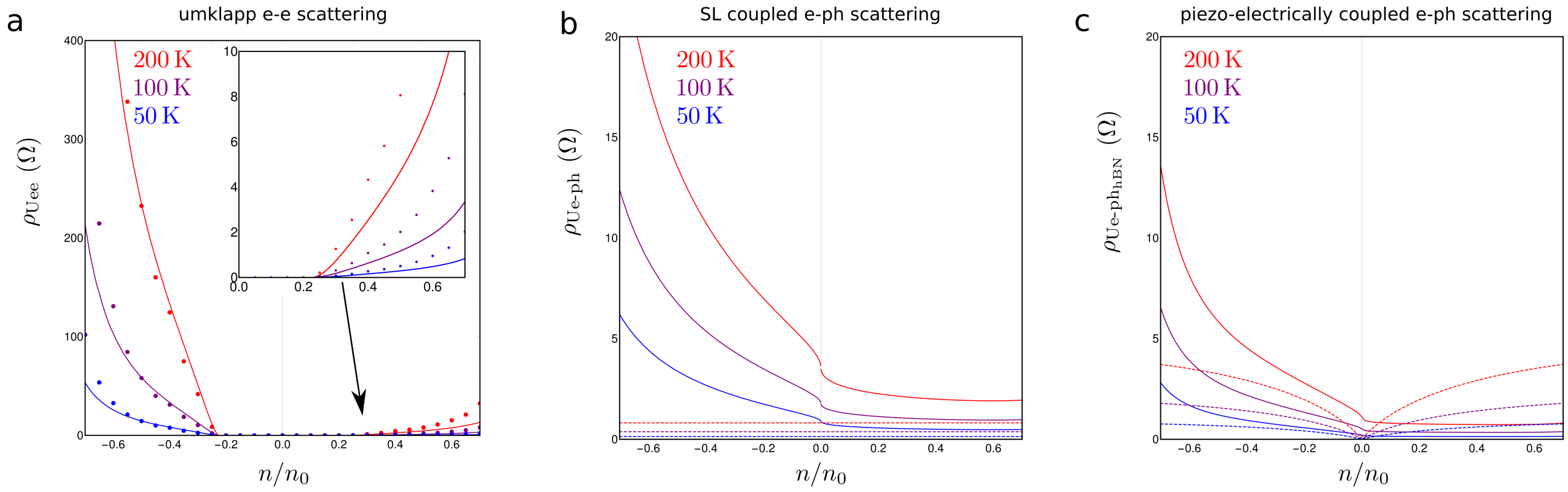

FIG. S5: **(a)** The calculated Uee contribution to the resistivity ($\rho_{\text{Uee}}$), calculated either using the method described in the main text (solid lines) or the method described in supplementary section S3 B 2 (dots). The inset displays the same data on a zoomed scale $n > 0$. **(b)** The resistivity, $\rho_{\text{Ue-ph}}$ , from SL umklapp scattering on in-plane phonons (solid lines, calculated using Eq. (S.32)) or out-of-plane phonons, $\rho_{\text{Ue-zph}}$, (dashed lines, calculated using Eq. (S.32)). **(c)** The resistivity, $\rho_{\text{Ue-ph}_{\text{hBN}}}$, from umklapp scattering on piezo-electrically coupled hBN phonons (solid lines, calculated using Eq. (S.41)) or normal scattering on piezo-electrically coupled hBN phonons, $\rho_{\text{e-ph}_{\text{hBN}}}$, (dashed lines, calculated using Eq. (S.39)). All curves are calculated using the SL strength parameters, $U_i$, given in the main text, $\delta = 1.8\%$, $\theta = 0°$, corresponding to $\lambda \approx 14$ nm and for temperatures $T = 50$, 100, and 200 K. We use $N_l = 10$ for (c). Note that a much higher resistivity scale is used in (a).

Figure S5(a) compares the Uee contribution to resistivity calculated either using the perturbation theory method described in the main text (solid lines), or, using the numerical approach described in supplementary section S3 B 2 which accounts for the SL reconstruction of the bandstructure and wavefunction. We note the good agreement between the two calculation methods.

The solid lines in Fig. S5(b) and (c) display $\rho_{\text{Ue-ph}}$ (calculated using Eq. (S.32)), and $\rho_{\text{Ue-ph}_{\text{hBN}}}$ (calculated using Eq. (S.41) for graphene on a 10 layer hBN slab, $N_l = 10$) respectively. The resistivity produced by both these scattering mechanisms is strongly asymmetric between n- and p-type doping, similar to the measured excess resistivity. However, the magnitude of the calculated resistivity is more than an order of magnitude smaller than the calculated Uee contribution to resistivity, Fig. S5(a), and the measured excess resistivity present in the main text. Moreover these resistivity contributions scale proportionally to the temperature (for $T \gg c_\nu g$) and scale quadratically with the SL period, in contrast to the measured trends displayed in Fig. 3 of the main text. Because of this, we neglect these contribution to the resistivity in the main text.

The dashed lines in Fig. S5(b) display the resistivity, $\rho_{\text{Ue-zph}}$ produced by the SL coupling with the out-of-plane modes (calculated using Eq. (S.33)). This produces a very small addition to the resistivity which is approximately independent of the doping in the flake since out-of-plane acoustic phonons are approximately non-dispersive for wavevectors $q \lesssim g$.

The dashed lines in Fig. S5(c) display the resistivity, $\rho_{\text{e-ph}_{\text{hBN}}}$, produced by normal scattering with piezo-electrically coupled hBN phonons (calculated using Eq. (S.39) for graphene on a 10 layer hBN slab). Note that this scattering mechanism produces an electron-hole symmetric resistivity that does not depend on the graphene-hBN alignment as it does not involve the moiré SL. Also note that the resistivity produced by both normal and umklapp scattering on piezo-electrically coupled hBN phonons is not significantly increased if the thickness of the hBN slab is increased ($N_l > 10$), due to the weaker coupling (Eq. (S.36)) to the further hBN layers.

---

[1] Mayorov, A. S., Gorbachev, R. V., Morozov, S. V., Britnell, L., Jalil, R., Ponomarenko, L. A., Blake, P., Novoselov, K. S., Watanabe, K., Taniguchi, T. & Geim, A. K., Micrometer-Scale Ballistic Transport in Encapsulated Graphene at Room Temperature. *Nano Lett.* **11**, 2396 (2011).

[2] Kretinin, A. V., Cao, Y., Tu, J. S., Yu, G. L., Jalil, R., Novoselov, K. S., Haigh, S. J., Gholinia, A., Mishchenko, A., Lozada, M., Georgiou, T., Woods, C. R., Withers, F., Blake, P., Eda, G., Wirsig, A., Hucho, C., Watanabe, K., Taniguchi, T , Geim, A. K. & Gorbachev, R. V. Electronic Properties of Graphene Encapsulated with Different Two-Dimensional Atomic Crystals. *Nano Lett.* **14**, 3270 (2014).

[3] M. Yankowitz, J. Xue, D. Cormode, J. D. Sanchez-Yamagishi, K. Watanabe, T. Taniguchi, P. Jarillo-Herrero, P. Jacquod & B. J. LeRoy. Emergence of superlattice Dirac points in graphene on hexagonal boron nitride. *Nat. Phys.* **8**, 382 (2012).

[4] Woods, C. R., Britnell, L., Eckmann, A., Ma, R. S., Lu, J. C., Guo, H. M., Lin, X., Yu, G. L., Cao, Y., Gorbachev, R. V., Kretinin, A. V., Park, J., Ponomarenko, L. A., Katsnelson, M. I., Gornostyrev, Yu. N., Watanabe, K., Taniguchi, T., Casiraghi, C., Gao, H-J., Geim, A. K. & Novoselov, K. S, Commensurateincommensurate transition in graphene on hexagonal boron nitride. *Nat. Phys.* **10**, 451 (2014).

[5] Wang, L., Meric, I., Huang, P. Y., Gao, Q., Gao, Y., Tran, H., Taniguchi, T., Watanabe, K., Campos, L. M., Muller, D. A., Guo, J., Kim, P., Hone, J., Shepard, K. L. & Dean, C. R. One-Dimensional Electrical Contact to a Two-Dimensional Material. *Science* **342**, 614 (2013).

[6] Ben Shalom, M., Zhu, M. J., Fal'ko, V. I., Mishchenko, A., Kretinin, A. V., Novoselov, K. S., Woods,C. R., Watanabe, K., Taniguchi, T., Geim, A. K. & Prance, J. R. Quantum oscillations of the critical current and high-field superconducting proximity in ballisticgraphene. *Nat. Phys.* **12**, 318 (2015).

[7] Krishna Kumar, R., Chen, X., Auton, G. H., Mishchenko, A., Bandurin, D. A., Morozov, S. V., Cao, Y., Khestanova, E., Ben Shalom, M., Kretinin, A. V., Novoselov, K. S., Eaves, L., Grigorieva, I. V., Ponomarenko, L. A., Fal'ko, V. I. & Geim, A. K. High-temperature quantum oscillations caused by recurring Bloch states in graphene superlattices. *Science* **357**, 181 (2017).

[8] Wallbank, J. R., Patel, A. A., Mucha-Kruczyński, M., Geim, A. K. & V. I. Falko. Generic miniband structure of graphene on a hexagonal substrate. *Phys. Rev. B* **87**, 245408 (2013).

[9] Ponomarenko, L. A., Gorbachev, R. V., Yu, G. L., Elias, D. C., Jalil, R., Patel, A. A., Mishchenko, A, Mayorov, A. S., Woods, C. R., Wallbank, J. R., Mucha-Kruczynski, M., Piot, B. A., Potemski, M., Grigorieva, I. V., Novoselov, K. S., Guinea, F., Fal'ko, V. I. & Geim A. K. Cloning of Dirac fermions in graphene superlattices. Nature **497**, 594 (2013).

[10] Dean, C. R., Wang, L., Maher, P., Forsythe, C., Ghahari, F., Gao, Y., Katoch, J., Ishigami, M., Moon, P., Koshino, M., Taniguchi, T., Watanabe, K., Shepard, K. L., Hone, J. & Kim, P.. Hofstadters butterfly and the fractal quantum Hall effect in moiré superlattices. Nature **497**, 598 (2013).

[11] Hunt, B., Sanchez-Yamagishi, J. D., Young, A. F., Yankowitz, M., LeRoy, B. J., Watanabe, K., Taniguchi, T., Moon, P., Koshino, M., Jarillo-Herrero, P. & Ashoori, R. C.. Massive Dirac Fermions and Hofstadter Butterfly in a van der Waals Heterostructure. Science **340**, 1427 (2013).

[12] Wallbank, J. R., Mucha-Kruczyński, M., Chen, X. & Falko, V I. Moiré superlattice effects in graphene/boron-nitride van der Waals heterostructures. *Ann. Phys.* **527**, 359 (2015).

[13] Lee, M., Wallbank, J. R., Gallagher, P., Watanabe, K., Taniguchi, T., Fal'ko, V. I., Goldhaber-Gordon, D. Ballistic

miniband conduction in a graphene superlattice. *Science* **353**, 1562 (2016).

[14] L. D. Landau and E. M. Lifshitz, Physical Kinetics (Pergamon Press Ltd., Oxford, 1981), Vol. 10, Sec. 79.

[15] J. M. Ziman, Electrons and Phonons: The Theory of Transport Phenomena in Solids (Oxfor University Press, 1960) Sec. 9.14.

[16] Cosma, D. A., Wallbank, J. R., Cheianov, V. & Fal'ko, V. I.. Moiré pattern as a magnifying glass for strain and dislocations in van der Waals heterostructures. *Faraday Discuss* **173**, 137 (2014).

[17] Lazzeri, M., Attaccalite, C., Wirtz, L. & Mauri, F. Impact of the electron-electron correlation on phonon dispersion: Failure of LDA and GGA DFT functionals in graphene and graphite. *Phys. Rev. B* **78**, 081406(R) (2008).

[18] Serrano, J., Bosak,A., Arenal, R., Krisch, M., Watanabe, K., Taniguchi, T., Kanda, H., Rubio, A. & Wirtz, L., Vibrational Properties of Hexagonal Boron Nitride: Inelastic X-Ray Scattering and Ab Initio Calculations. *Phys. Rev. Lett.* **98**, 095503 (2007).

[19] Mostaani, E., Drummond, N. D. & Fal'ko, V. I. Quantum Monte Carlo Calculation of the Binding Energy of Bilayer Graphene. *Phys. Rev. Lett.* **115**,115501 (2015).

[20] Duerloo, K.-A. N., Ong, M. T. & Reed, E. J. Intrinsic Piezoelectricity in Two-Dimensional Materials. *J. Phys. Chem. Lett.* **3**, 2871 (2012).

[21] Michel, K. H., Çakr, D., Sevik, C. & Peeters, F. M. Piezoelectricity in two-dimensional materials: Comparative study between lattice dynamics and ab initio calculations. *Phys. Rev. B* **95**, 125415 (2017).

[22] Rostami, H., Guinea, F., Polini, M. & Roldán, R. Piezoelectricity and valley Chern number in inhomogeneous hexagonal 2D crystals. arXiv:1707.03769

[23] Slotman, G. J. , de Wijs, G. A., Fasolino, A. & Katsnelson, M. I. Phonons and electron-phonon coupling in graphene-h-BN heterostructures. *Ann. Phys.* **526**, 381 (2014).

[24] Nemanich, R. J., Solin, S. A. & Martins, R. M. Light scattering study of boron nitride microcrystals *Phys. Rev. B* **23**, 6348 (1981).